\documentclass[twocolumn]{aastex631} 

\usepackage{amsmath}	
\usepackage{amssymb}	
\usepackage{wasysym}
\usepackage{multirow}
\usepackage[flushleft]{threeparttable}
\usepackage{booktabs}

\renewcommand{\arraystretch}{1.25} 

\shorttitle{Emulating Model Grids with Normalizing Flows}
\shortauthors{Hon et al.}

\graphicspath{{./}{figures/}}

\begin{document}

\title{Flow-Based Generative Emulation of Grids of Stellar Evolutionary Models}

\correspondingauthor{Marc Hon}
\email{mtyhon@mit.edu}

\author[0000-0003-2400-6960]{Marc Hon}
\affiliation{Kavli Institute for Astrophysics and Space Research, Massachusetts Institute of Technology,
77 Massachusetts Avenue, Cambridge, MA 02139, USA}
\affiliation{Institute for Astronomy, University of Hawai`i,
2680 Woodlawn Drive, Honolulu, HI 96822, USA}

\author[0000-0003-3020-4437]{Yaguang Li}
\affiliation{Institute for Astronomy, University of Hawai`i,
2680 Woodlawn Drive, Honolulu, HI 96822, USA}

\author[0000-0001-7664-648X]{Joel Ong}
\affiliation{Institute for Astronomy, University of Hawai`i,
2680 Woodlawn Drive, Honolulu, HI 96822, USA}

\begin{abstract}
We present a flow-based generative approach to emulate grids of stellar evolutionary models. By interpreting the input parameters and output properties of these models as multi-dimensional probability distributions, we train conditional normalizing flows to learn and predict the complex relationships between grid inputs and outputs in the form of conditional joint distributions. Leveraging the expressive power and versatility of these flows, we showcase their ability to emulate a variety of evolutionary tracks and isochrones across a continuous range of input parameters. In addition, we describe a simple Bayesian approach for estimating stellar parameters using these flows and demonstrate its application to asteroseismic datasets of red giants observed by the \textit{Kepler} mission. By applying this approach to red giants in open clusters NGC 6791 and NGC 6819, we illustrate how large age uncertainties can arise when fitting only to global asteroseismic and spectroscopic parameters without prior information on initial helium abundances and mixing length parameter values. We also conduct inference using the flow at a large scale by determining revised estimates of masses and radii for 15,388 field red giants. These estimates show improved agreement with results from existing grid-based modelling, reveal distinct population-level features in the red clump, and suggest that the masses of \textit{Kepler} red giants previously determined using the corrected asteroseismic scaling relations have been overestimated by $5-10\%$.
\end{abstract}



\section{Introduction} \label{sec:intro} 

Stellar models\footnote{The term `model' in this work refers only to simulated physical profiles of stars. This is to disambiguate from its meaning when used in the context of machine learning.} form a cornerstone of contemporary astrophysics by enabling a direct comparison between observations and theory. A wide variety of variables dictate the evolution of stellar models, including fundamental input parameters (e.g., initial masses and chemical composition), parameters defining the prescription of input physics, the treatment of stellar convection \citep{Kupka_2017}, and the presence of stellar rotation (e.g., \citealt{Ekstrom_2012}). Conversely, stellar models predict a broad range of observables such as stellar luminosities, effective temperatures, chemical abundances, and pulsation periods. Such observables are fundamental to identifying which model best aligns with observed data, enabling properties that can only be inferred from models \textemdash{} such as stellar ages \textemdash{} to be determined \citep{Soderblom_2010}.

The broad diversity in both input and output parameters for grids of stellar models commonly poses a challenge for using grids to study the evolution and astrophysical properties of stars. A wide range of input parameters results in many free parameters dictating the evolution of models, resulting in the curse of dimensionality that makes the systematic exploration of models across the grid computationally infeasible.
The outputs of stellar models, which are often observables, can also be high-dimensional, with complex and non-linear dependencies with input parameters. This complexity in inferring observables from grids of models has naturally led to the adoption of machine learning approaches in forward modelling tasks in grids of stellar models (e.g., \citealt{Verma_2016, Bai_2019, Bellinger_2020, Hon_2020,Garraffo_2021, Lyttle_2021, Mombarg_2021, Scutt_2023, Maltsev_2024, Panda_2024}). Many approaches frame the forward modelling task as a one-to-one prediction task, in which a singular model \textemdash{} one whose predicted observables align most closely with the observed data \textemdash{} is the emulated output from the machine learning algorithm. However, there often exists degeneracies within the grid, whereby multiple combinations of input parameters can produce the same observable. One-to-one predictions do not handle these scenarios effectively.
\citet{Li_2022} introduced a one-to-many prediction approach for forward modelling task by emulating regions of the grid using Gaussian process regressors. These regressors generate continuous functions that map inputs to observables, ensuring smooth correlations between inputs and observables across the grid that is expected in stellar evolution. A notable advantage of this emulation method is that forward modelling is conducted by sampling from the emulated grid, because it leverages statistical methods for a more transparent measurement of stellar model properties, in contrast to the direct prediction of properties from black-box algorithms like neural networks.

Building on these insights, we present the use of normalizing flows as a highly flexible and expressive global emulator of grids of stellar models.
Normalizing flows are a class of probabilistic machine learning techniques that learn invertible transformations between simple base distributions and more complex distributions \citep{NICE, RealNVP}. Because such transformations are typically learned by training deep neural networks, normalizing flows can conduct the density estimation and sampling of diverse and elaborate high-dimensional distributions. 
As a result, these approaches have been applied in complex generative tasks and in modern 
probabilistic inference approaches including high-dimensional statistical modelling (e.g., \citealt{Ting_2022, Van-Lane_2023}) and likelihood-free inference (e.g., \citealt{Wang_2023}). Here, we apply normalizing flows to grids of stellar evolutionary models to emulate the input and output properties of the grids as high-dimensional distributions.

We demonstrate three main capabilities of normalizing flows for this task. The first is the emulation of stellar evolutionary tracks and isochrones, which highlights the complex relations between fundamental properties of stellar evolutionary models and their outputs. Visualizing how evolutionary tracks vary with initial stellar parameters is fundamental to our understanding of stellar evolution and for comparative analyses between stars (e.g., \citealt{White_2011, Gai_2017}). Due to the high dimensionality of the model grid, visualizing variations in input parameters is typically conducted one parameter at a time and at discrete intervals. This approach leads to limited coverage of the input parameter space and induces large uncertainties when attempting to pinpoint the properties of a specific star. We show that normalizing flows can mitigation this limitation by providing expressive and smooth high-dimensional interpolations within a grid of models.

The second capability involves examining correlations in high dimensions, demonstrated using red giant stars in the open clusters NGC 6791 and NGC 6819 that were observed by the \textit{Kepler} mission \citep{Borucki_2010}. The co-eval nature of stellar clusters provides a unique testbed for the input physics to stellar models, and this has resulted in a broad range of derived cluster ages across literature from various model fitting approaches to asteroseismic data (e.g., \citealt{Basu_2011, Kallinger_2018, Mckeever_2019, Li_2022}), isochrones (e.g., \citealt{Stetson_2003, Jeffries_2013}), eclipsing binaries (e.g., \citealt{Brewer_2016}), and white dwarfs (e.g., \citealt{Garcia_2010}). Various input physics and parameters to grids of stellar models are adopted across these approaches, and thus it is particularly insightful to understand how different grid input parameters affect age, which may explain the spread of modeled ages in literature. Given that normalizing flows are useful for examining high-dimensional correlations, we utilize the flows to investigate the spread of ages for the two open clusters.

As a generative approach, normalizing flows provides an alternative approach to grid-based inference by the sampling of emulated evolutionary tracks. We demonstrate this capability by estimating the fundamental properties of an ensemble of \textit{Kepler} field red giants, whose population-level distributions of masses and radii are critical for Galactic archeology studies (e.g., \citealt{Pinsonneault_2018, Silva-Aguirre_2018, Miglio_2021, Anders_2023}). Yet, existing grid-based modelling approaches have been applied to only a subset of the stars in the \textit{Kepler} field, whereas more extensive studies of red giants in the \textit{Kepler} field to date have utilized asteroseismic scaling relations adjusted by a correction factor computed by simple grid interpolation \citep{Sharma_2016, Yu_2018}. The masses using the corrected scaling relations, however, result in a population-level distribution that disagrees with synthetic populations of the Milky Way \citep{Sharma_2016, Sharma_2019}, which motivates our re-examination of the masses and radii of the \textit{Kepler} red giants in this work.

Notably, our use of flows in this work parallels the study by \citet{Ksoll_2020}, which used the functionally similar Invertible Neural Networks to determine stellar parameters based on photometric observations. However, the novelty of our study lies in our emphasis towards the smooth interpolation of conditioning variables in normalizing flows. 
A key difference, as described in Section \ref{section:flows}, is our use of input grid parameters rather than observables as conditioning variables. This is a deliberate design choice that demonstrate the flexibility of normalizing flows in inferring stellar parameters as a many-to-many prediction algorithm.

\begin{figure}
    \centering
	\includegraphics[width=1.\columnwidth]{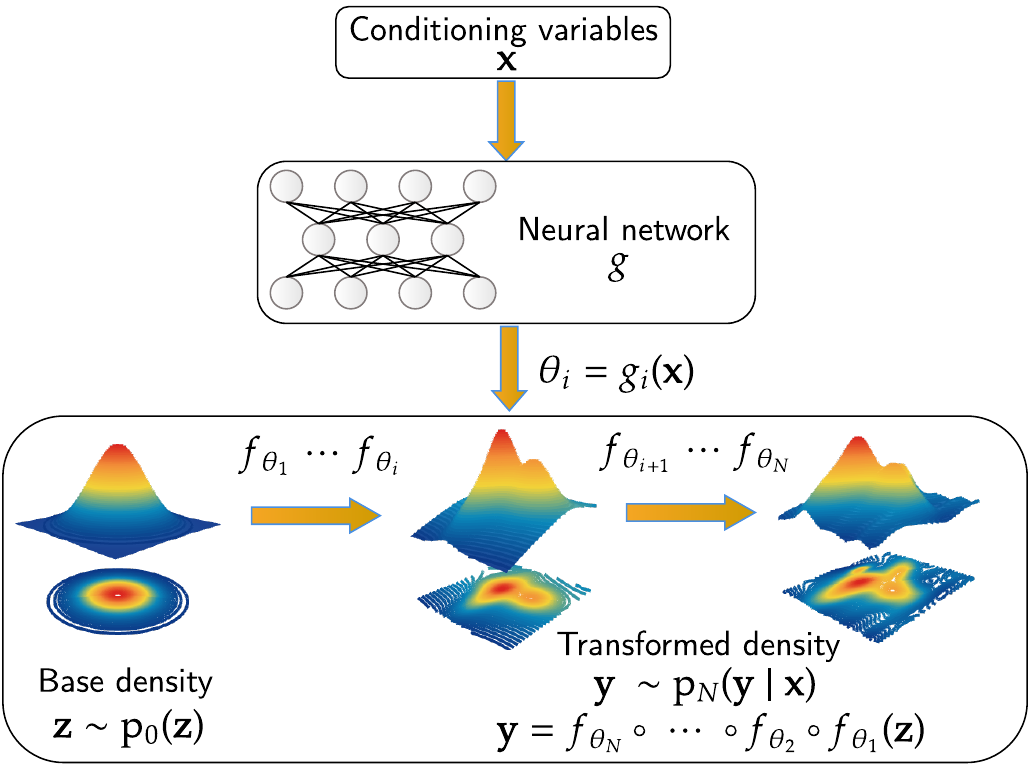}
    \caption{A schematic of conditional normalizing flows. The flow maps random variates $\mathbf{z}$ from a base probability density $\mathbf{p_0(\mathbf{z})}$ (here shown as a 2D normal distribution) to another variable $\mathbf{y}$. The mapping occurs over a series of $N$ invertible transformations $f = f_{\theta_i}, i \in [1, 2, \cdots, N]$, where $\theta_i$ is learned by a neural network and is conditioned by contextual inputs $\mathbf{x}$. The probability density of $\mathbf{y}$ is subsequently conditioned by $\mathbf{x}$, such that $y_N \sim p_N(\mathbf{y}|\mathbf{x})$. In this work, the $\mathbf{y}$ corresponds to the output stellar properties of an evolutionary grid of models, while $\mathbf{x}$ corresponds to the input parameters of the grid.}
    \label{fig:normflowsdiagram}
\end{figure}

\begin{table*}[ht]
\centering
\renewcommand{\arraystretch}{1.2}
\begin{threeparttable}
\caption{Variables from MESA stellar models used for density estimation using the Conditional Normalizing Flow focused on subgiant and dwarf star models, CNF$_{\mathrm{dwarf}}$.}\label{table:mesa}
\begin{tabular}{cp{10cm}c} 
\toprule
\midrule
\multicolumn{3}{c}{Conditioning Input Parameters ($\mathbf{x}$)}\\
\midrule
\textbf{Parameter} & \textbf{Definition} & \textbf{Range} \\
\midrule
$M$ & Stellar mass$^1$ & $0.7\,M_{\odot}\leq M \leq 2.5\,M_{\odot}$\\
$\log_{10}Z$ & Initial metal fraction & $-4.934 \leq \log_{10}Z \leq -1.291$ \\
$Y$ & Initial helium fraction & $0.23 \leq Y \leq 0.37$ \\
$\alpha$ & Mixing length parameter & $1.0 \leq \alpha \leq 2.7$ \\
\multirow{2}{*}{$\widehat{f_{\mathrm{ov, env}}}$} & Min-max normalized$^2$ $f_{\mathrm{ov, env}}$, where $f_{\mathrm{ov, env}}$ is the efficiency parameter of convective overshoot in stellar envelope & \multirow{2}{*}{$2\times10^{-6} \leq f_{\mathrm{ov, env}} \leq 2\times10^{-2}$} \\
\multirow{2}{*}{$\widehat{f_{\mathrm{ov, core}}}$} & Min-max normalized$^2$ $f_{\mathrm{ov, core}}$, where $f_{\mathrm{ov, core}}$ is the efficiency parameter of convective overshoot in stellar core & \multirow{2}{*}{$3\times10^{-6} \leq f_{\mathrm{ov, core}} \leq 3\times10^{-2}$} \\
\midrule
\bottomrule
\multicolumn{3}{c}{Output Properties ($\mathbf{y}$)} \\ 
\midrule
\textbf{Property} & \textbf{Definition} & \textbf{Range} \\
\midrule
$\widehat{T_{\mathrm{eff}}}$ & $z$-normalized$^3$ $\log_{10}T_{\mathrm{eff}}$, where $T_{\mathrm{eff}}$ is stellar effective temperature &  $3344\,$K $\leq T_{\mathrm{eff}} \leq 19252\,$K  \\
\multirow{2}{*}{$\widehat{\nu_{\mathrm{max}}}$} & $\log_{10}\nu_{\mathrm{max}}$, where $\nu_{\mathrm{max}}$ is the frequency at maximum power estimated by MESA& \multirow{2}{*}{$300\,\mu$Hz $\leq \nu_{\mathrm{max}} \leq 6684\,\mu$Hz}\\
\multirow{4}{*}{$\widehat{\Delta\nu}$}  & $\log_{10}\Delta\nu$, where $\Delta\nu$ is the large frequency separation value computed using a weighted linear fit to $l=0$ mode frequencies around $\nu_{\mathrm{max}}$. The weights are determined by fitting a Gaussian envelope$^4$ centered about $\nu_{\mathrm{max}}$ to the $l=0$ mode frequencies  & \multirow{4}{*}{$17.94\,\mu$Hz $\leq \Delta\nu \leq 232.48\,\mu$Hz} \\
\multirow{2}{*}{$\epsilon$}  & Dimensionless phase offset of the asymptotic relation of p-mode frequencies. Determined as the intercept of the weighted linear fit to $\Delta\nu$. & \multirow{2}{*}{$0.43$ $\leq \epsilon \leq 2.41$} \\
\multirow{4}{*}{$\widehat{\delta\nu_{01}}$}  & z-normalized$^2$ $\delta\nu_{01}$, where $\delta\nu_{01}$ is the frequency separation between $l=0$ and $l=1$ frequencies. This quantity is computed as $\delta\nu_{01} = \frac{1}{2}(\nu_{n, l=0} +\nu_{n+1, l=0}) -\nu_{n, l=1}$, averaged across radial orders ($n$) using weights determined by fitting a Gaussian envelope$^4$ centered about $\nu_{\mathrm{max}}$ to the $l=1$ mode frequencies & \multirow{4}{*}{$-4.42\,\mu$Hz $\leq \delta\nu_{01} \leq 13.42\,\mu$Hz} \\
\multirow{4}{*}{$\widehat{\delta\nu_{02}}$}  & $\log_{10}\delta\nu_{02}$, where $\delta\nu_{02}$ is the small frequency separation value computed as $\delta\nu_{02} = \nu_{n, l=0} +\nu_{n-1, l=2}$. This quantity is computed as $\nu_{0} -\nu_{2}$, averaged across radial orders using weights determined by fitting a Gaussian envelope$^4$ centered about $\nu_{\mathrm{max}}$ to the $l=2$ mode frequencies & \multirow{4}{*}{$0.31\,\mu$Hz $\leq \delta\nu_{02} \leq 20.92\,\mu$Hz} \\
\multirow{4}{*}{$\widehat{\delta\nu_{03}}$}  & $\log_{10}\delta\nu_{03} + 3$, where $\delta\nu_{03}$ is the frequency separation between $l=0$ and $l=3$ frequencies. This quantity is computed as $\delta\nu_{01} = \frac{1}{2}(\nu_{n, l=0} +\nu_{n+1, l=0}) -\nu_{n-1, l=3}$, averaged across radial orders using weights determined by fitting a Gaussian envelope$^4$ centered about $\nu_{\mathrm{max}}$ to the $l=3$ mode frequencies & \multirow{4}{*}{$-1.9\,\mu$Hz $\leq \delta\nu_{03} \leq 41.0\,\mu$Hz} \\
$\hat{R}$  & $\log_{10}R$, where $R$ is stellar radius in units of $R_{\odot}$ &  $0.61\,R_{\odot}\leq R \leq 5.08\,R_{\odot}$\\
$\hat{\tau}$ & $\log_{10}\tau$, where $\tau$ is stellar age in units of Gyr &  $0.0013\,$Gyr $\leq\tau\leq100\,$Gyr \\
\midrule
\bottomrule
\end{tabular}
\begin{tablenotes}
\item[1] Represents both the initial and the current mass of the model. 
\item[2] The quantity is squeezed into an interval $\in[-1,1]$ by first subtracting its minimum value across the dataset and then dividing by the difference between the maximum and minimum values over the dataset. 
\item[3] The quantity is first subtracted by its mean value over the dataset, then divided by its standard deviation over the dataset.
\item[4] The width of the envelope is determined as $W = \nu_{\mathrm{max}}^k \cdot e^b$, where $k=0.9638, b=-1.7145$ based on parametric fits to the widths of \textit{Kepler} asteroseismic targets \citep{Li_2020}.
\end{tablenotes}
\end{threeparttable}
\end{table*} 

\begin{table*}[ht]
\centering
\begin{threeparttable}
\caption{Variables from the grid of models accompanying AsfGrid used for training the CNF$_{\mathrm{asfgrid}}$ Conditional Normalizing Flow. }\label{table:asfgrid}
\begin{tabular}{cp{10cm}c} 
\toprule
\midrule
\multicolumn{3}{c}{Conditioning Input Parameters ($\mathbf{x}$)}\\
\midrule
\textbf{Parameter} & \textbf{Definition} & \textbf{Range} \\
\midrule
$M$ & Stellar mass$^1$ & $0.6\,M_{\odot}\leq M \leq 5.5\,M_{\odot}$\\
\multirow{2}{*}{$[\text{Fe/H}]$} & Metallicity, determined from the grid as $\log_{10}Z/Z_{\odot}$, where $Z_{\odot}=0.019$ following \citet{Sharma_2016} & \multirow{2}{*}{$-3\, \text{dex} \leq [\text{Fe/H}] \leq 0.4\,\text{dex}$}
\\
\multirow{3}{*}{$E$} & Evolutionary state, defined as $-1$ for lower giant branch stars with $\nu_{\mathrm{max}}\geq300\,\mu$Hz; 0 for hydrogen shell-burning red giant stars with $10\mu$Hz$\,\leq\nu_{\mathrm{max}}<300\,\mu$Hz; 1 for hydrogen shell-burning giant stars with $\nu_{\mathrm{max}}<10\,\mu$Hz; 2 for helium-core burning stars & \multirow{3}{*}{$E \in [-1, 0, 1, 2]$}\\
\midrule
\bottomrule
\multicolumn{3}{c}{Output Properties ($\mathbf{y}$)} \\ 
\midrule
\textbf{Property} & \textbf{Definition} & \textbf{Range} \\
\midrule
$\widehat{T_{\mathrm{eff}}}$ & $\log_{10}T_{\mathrm{eff}}$, where $T_{\mathrm{eff}}$ is stellar effective temperature &  $2910\,$K $\leq T_{\mathrm{eff}} \leq 22800\,$K  \\
\multirow{2}{*}{$\widehat{\Delta\nu}$}  & $\log_{10}\Delta\nu$, where $\Delta\nu$ is AsfGrid's large frequency separation value computed using radial mode frequencies using GYRE \citep{Townsend_2013} & \multirow{2}{*}{$0.12\,\mu$Hz $\leq \Delta\nu \leq 77.42\,\mu$Hz} \\
\multirow{2}{*}{$\widehat{\nu_{\mathrm{max}}}$} & $\log_{10}\nu_{\mathrm{max}}$, where $\nu_{\mathrm{max}}$ is the frequency at maximum power estimated by MESA as reported by the grid accompanying AsfGrid& \multirow{2}{*}{$0.05\,\mu$Hz $\leq \nu_{\mathrm{max}} \leq 1000\,\mu$Hz}\\
$\hat{R}$  & $\log_{10}R$, where $R$ is stellar radius in units of $R_{\odot}$ &  $1.5\,R_{\odot}\leq R \leq 193\,R_{\odot}$\\
$\hat{\tau}$ & $\log_{10}\tau$, where $\tau$ is stellar age in units of Gyr &  $0.05\,$Gyr $\leq\tau\leq53\,$Gyr \\
\midrule
\bottomrule
\end{tabular}
\begin{tablenotes}
\item[1] Represents both the initial and the current mass of the model. 
\end{tablenotes}
\end{threeparttable}
\end{table*} 

\section{Methods}
\subsection{Normalizing Flows}\label{section:flows}
The central idea of a normalizing flow is to map a random latent variable $\mathbf{z}$, which is distributed following a simple probability density $p_0(\mathbf{z})$ (typically a multivariate normal), to another variable $\mathbf{y}=f(\mathbf{z})$ that is distributed following a more complex probability density $p_N(\mathbf{y})$. 
The transformation $f = f_{\theta_N} \circ \dots f_{\theta_2} \circ f_{\theta_1}$ is a series of functions parameterized by $\theta \in \theta_1, \theta_2, \cdots \theta_N$, whose values are learned by deep neural networks. Such a transformation describes how samples $\mathbf{z}\sim p_0(\mathbf{z})$ are mapped to $\mathbf{y}\sim p_N(\mathbf{y})$ following the change of variables formula:
\begin{align}\label{eqn:1}
    p_0(\mathbf{z}) &= p_N(\mathbf{y}) \cdot \left| \det \left( \frac{d\mathbf{y}}{d\mathbf{z}} \right) \right| \notag \\
    & = p_N(f(\mathbf{z})) \cdot \left| \det \left( J_f(\mathbf{z}) \right) \right|,
\end{align}
\noindent where $J_f$ denotes the Jacobian of the transformation $f$. Training a normalizing flow to perform density estimation involves the optimization of $\theta$ to maximize the (log-)likelihood of observed data under the probability density described by the flow:
\begin{equation}\label{eqn:log-likelihood}
\log p_N(\mathbf{y}) = \log p_0(f^{-1}(\mathbf{y})) + \log \left| \det \left( \frac{\partial f^{-1}(\mathbf{y})}{\partial \mathbf{y}} \right) \right|
\end{equation}
From Equation \ref{eqn:log-likelihood}, a normalizing flow evidently requires the transformation $f$ to have the following properties:

\begin{itemize}
    \item \textbf{Differentiability:} This property enables the factor $\left| \det \left( J_f(\mathbf{z}) \right) \right|$ in Equation \ref{eqn:1} to be computed, which ensures that $p_N$ integrates to unity and remains a valid probability density under the transformation. Additionally, the differentiability of the transformation permits efficient gradient-based optimization methods to be used when training the normalizing flow.
    \item \textbf{Invertibility:} This property ensures a one-to-one correspondence between $\mathbf{y}$ and $\mathbf{z}$ under the specified transformations. Consequently, the exact likelihood of samples $\mathbf{y}$ can be computed. The condition of invertibility also necessitates that the dimensionality of both $p_0$ and $p_N$ are equal.
\end{itemize}
In literature, there are various classes of transformations that inherit both of these properties. In this work, we use Autoregressive Neural Spline Flows \citep{Durkan_2019}, which are Masked Autoregressive Transforms \citep{Papamakarios_2017} combined with Monotonic Rational Quadratic Spline coupling transforms \citep{Durkan_2019} as implemented by the \texttt{zuko}\footnote{\url{https://zuko.readthedocs.io}} library version 0.3.2 \citep{Zuko}. A brief explanation of these transforms and their implementation in this work are provided in Appendix \ref{appendix:flows}.

Critically, the deep neural networks involved in the estimation of $\theta$ can be conditioned on contextual input variables $\mathbf{x}$, such that $\theta\rightarrow\theta(\mathbf{x}$). Consequently, the generative model is modified into a Conditional Normalizing Flow (CNF, \citealt{Winkler_2019}) and yields conditional probability densities $p_N(\mathbf{y} | \mathbf{x})$, as illustrated in Figure \ref{fig:normflowsdiagram}. By incorporating conditioning variables, a CNF leverages the expressiveness of neural networks to estimate conditional and marginal probability distributions across the range of each grid's input parameters. This is performed by optimizing the CNF's neural network to maximize the likelihood of the distribution described by the grid of models under the predicted conditional distribution from the CNF. 

\begin{figure*}
    \centering
	\includegraphics[width=2.0\columnwidth]{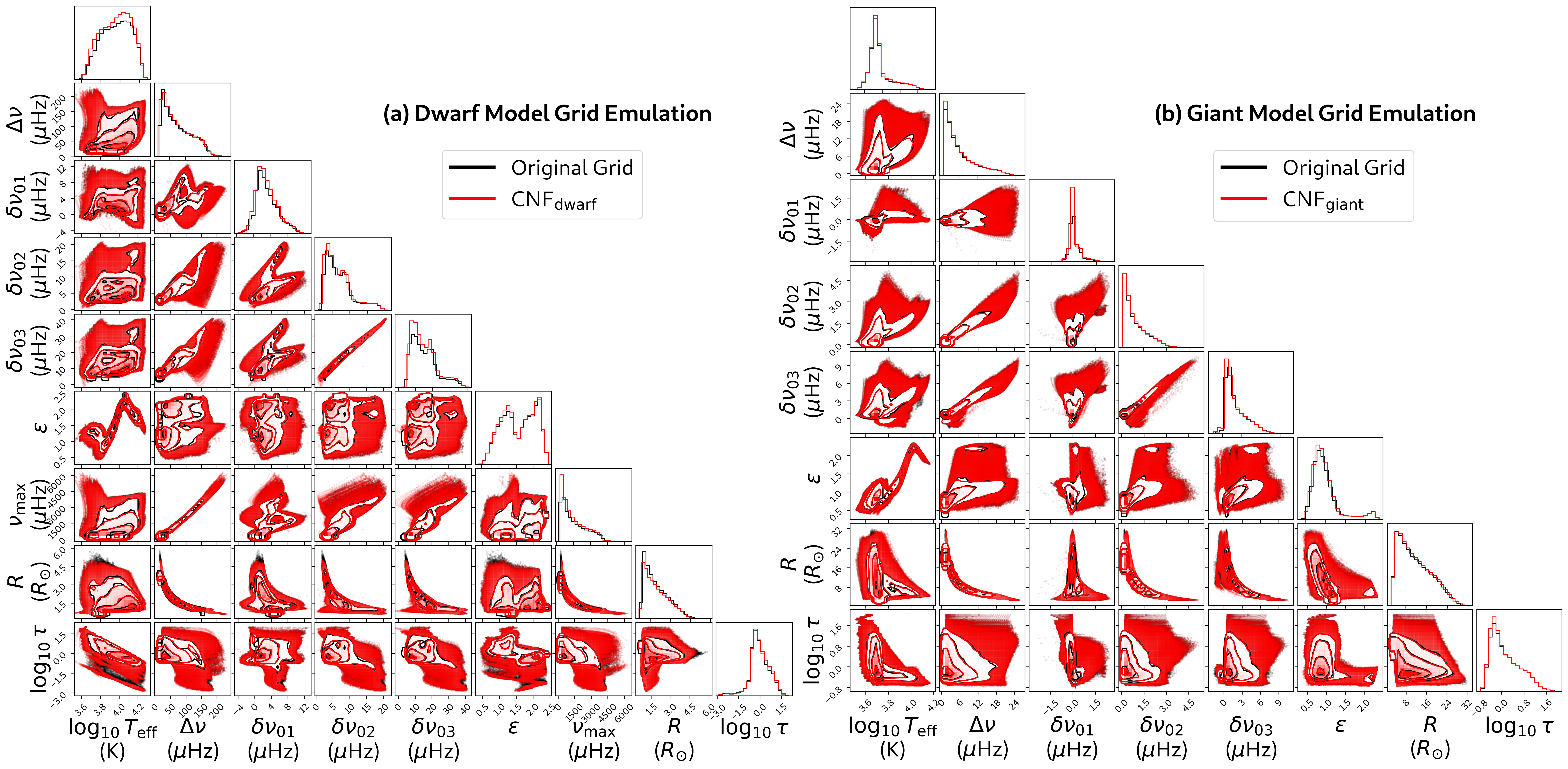}
    \caption{\textbf{(a)} Corner plot showing the original distribution (black) of output parameters $\mathbf{y}$ in the MESA grid of dwarf models and the emulated grid from CNF$_{\mathrm{dwarf}}$ in red. \textbf{(b)} Same as (a), except the comparison is between the MESA grid of giant models and the emulated grid from CNF$_{\mathrm{giant}}$. Descriptions of each parameter are listed in Table \ref{table:mesa}.}
    \label{fig:cornerplot}
\end{figure*}

\subsection{Grids of Evolutionary Models}\label{section:grid}

\subsubsection{MESA Grid of Dwarf Star Models} \label{grid_dwarfs}
We train a CNF on a grid of models generated using Modules for Experiments in Stellar Astrophysics (MESA, version r23.01.1;
\citealt{Paxton_2011, Paxton_2013, Paxton_2015, Paxton_2018, Paxton_2019}).
Our prescription for the grid of models closely follows that from \citet{Li_2023}. More specifically, we adopt the \citet{Asplund_2009} solar mixture, whereby $X_{\odot} = 0.7381, Y_{\odot} = 0.2485, Z_{\odot} = 0.0134$ and the metallicity [M/H] is correspondingly defined as $\text{[M/H]} = \log_{10}(Z/X) - \log_{10}(Z_{\odot}/X_{\odot})$. We select opacity tables in MESA following this choice of metal mixture, which are described by a combination of electron conduction opacities \citep{Cassisi_2007}, OPAL radiative opacities \citep{Iglesias_1993, Iglesias_1996}, low-temperature opacities \citep{Ferguson_2005} and data at the high-temperature Compton-scattering regime \citep{Buchler_1976}. We use the equation of state provided by MESA, which combines data from OPAL \citep{Rogers_2002}, SCVH \citep{Saumon_1995}, PTEH \citep{Pols_1995}, HELM \citep{Timmes_2000}, and PC \citep{Potekhin_2010}. We adopt nuclear reaction rates from JINA REACLIB database \citep{Cyburt_2010}, using a minimal set of elements specified in MESA's \texttt{basic.net}. Both atomic diffusion and gravitational settling are not included in the models. Convection is treated using the \citet{Henyey_1965} formalism, whose efficiency is controlled by a mixing length parameter $\alpha$. We define the boundary of convective regions following the
Schwarzschild criterion \citep{Schwarzschild_1958} and apply overshooting in both the envelope and core following the exponential scheme described by \citet{Herwig_2000}, with the corresponding efficiency parameters $f_{\mathrm{ov, env}}$ and $f_{\mathrm{ov, core}}$, respectively. For the treatment of 
surface boundary conditions, we adopt the grey model atmosphere in conjunction with the Eddington $T-\tau$ integration method \citep{Eddington_1926}. 
The models do not include the effects of stellar rotation, nor do they include mechanisms for mass loss.
We use GYRE (version 7.0, \citealt{Townsend_2013}) to calculate adiabatic frequencies from the structure profiles computed from MESA. Besides the frequencies of radial modes ($\ell=0$), we also calculate the frequencies of decoupled (pure p) dipolar ($\ell=1$) and quadrupolar modes ($\ell=2$) and octupolar modes ($\ell=3$) following the \citet{Ong_2020} decoupling approach.

The conditioning variable vector $\mathbf{x}$ comprises quantities over which evolutionary tracks in the grid are varied. These are $[M$, $\log_{10}Z$, $Y$, $\alpha$, $\widehat{f_{\mathrm{ov, env}}}$, $\widehat{f_{\mathrm{ov,core}}}]$, with 
the definitions of these input parameters presented in Table \ref{table:mesa}.
These input parameters are varied by Sobol sequence sampling \citep{Sobol_1967}, which is a quasi-random point generation scheme for homogeneously populating the hyperrectangle spanned by $\mathbf{x}$.  More specifically, we generate $2^{13}=$8,192 Sobol numbers within a six-dimensional unit hypercube and map them to the parameter ranges of each quantity in $\mathbf{x}$. As described by \citet{Bellinger_2016}, this scheme ensures a uniform coverage of the high-dimensional grid, which minimizes redundant information by preventing points from occupying similar regions in high-dimensional space. 
The number of samples represents a trade-off between computational cost and coverage of the parameter space, from which we find that using $2^{13}$ samples strikes an appropriate balance between these two factors. An illustration of the sampling coverage for $\mathbf{x}$ is presented in Appendix \ref{appendix:sobol}.

Each evolutionary track in the grid is evolved from the zero age main sequence up to pre-core helium ignition phase at the tip of the red giant branch. The CNF is trained to estimate an 9D probability density $p(\mathbf{y} | \mathbf{x})$, with $\mathbf{y} = [\widehat{T_{\mathrm{eff}}}, \widehat{\Delta\nu}, \widehat{\nu_{\mathrm{max}}}, \widehat{\delta\nu_{01}},  \widehat{\delta\nu_{02}}, \widehat{\delta\nu_{03}}, \epsilon, \hat{R}, \hat{\tau}]$ as outputs from the model (see Table \ref{table:mesa}) at each evolutionary timestamp along the grid. To focus on the density modelling of subgiant and dwarf star models within the grid, we select only models with $\nu_{\mathrm{max}}>300\mu$Hz for this particular CNF, resulting in a total of 443,874 models in this grid. We denote the CNF trained on this grid as $\text{CNF}_{\text{dwarf}}$.

\subsection{MESA Grid of Giant Star Models} \label{grid_giants}
The same base grid in Section \ref{grid_dwarfs} is adopted here, with two differences. First, we select only models with $\nu_{\mathrm{max}} \leq 300\mu$Hz to focus the density estimation task on models on the red giant branch, which results in a grid of 738,939 models. This $\nu_{\mathrm{max}}$ cutoff is specifically chosen to be near the Nyquist frequency of the Kepler 30-min sampling cadence at $\sim283\,\mu$Hz, below which exists the oscillation frequencies of most \textit{Kepler} red giants analyzed to date. Therefore, this CNF is tailored specifically for the emulation-based inference of \textit{Kepler} giants, which is described in Section \ref{section:open_clusters}. Compared to CNF$_{\mathrm{dwarf}}$, this giant star-specific CNF uses $\nu_{\mathrm{max}}$ as a conditioning variable instead of an output property to simplify selecting emulated models at a specific evolutionary stage. Therefore, we now have $\mathbf{x}=$ $[M$, $\log_{10}Z$, $Y$, $\alpha$, $\widehat{f_{\mathrm{ov, env}}}$, $\widehat{f_{\mathrm{ov,core}}}$, $ \widehat{\nu_{\mathrm{max}}}$$]$ and $\mathbf{y} = [\widehat{T_{\mathrm{eff}}}, \widehat{\Delta\nu}, \widehat{\delta\nu_{01}},  \widehat{\delta\nu_{02}}, \widehat{\delta\nu_{03}}, \epsilon, \hat{R}, \hat{\tau}]$. We denote the CNF trained on this grid as $\text{CNF}_{\text{giant}}$.

\subsubsection{AsfGrid Evolutionary Models} \label{asfgrid}
We train a CNF on a grid accompanying the publicly available Python module AsfGrid version 0.06 \citep{Sharma_2016, Stello_2022}. This is a grid of stellar evolutionary models used for ensemble asteroseismic studies of red giants (e.g., \citealt{Yu_2018, Zinn_2020, Zinn_2022, Li_2023}). It adopts a quasi-uniform sampling in $M$ and [Fe/H] across evolutionary tracks, as shown in Figure 1 in \citet{Stello_2022}. To specifically focus on the density modelling of red giant stars within the grid, we only select stellar models with $\nu_{\mathrm{max}}\leq1000\mu$Hz on the red giant branch (\texttt{evstate} = 1) and models in the red clump phase (\texttt{evstate} = 2) from the grid, resulting in a total of 1,071,210 models. The CNF is trained to estimate the conditional density $p(\mathbf{y}| \mathbf{x}$), where $\mathbf{y} = [\widehat{T_{\mathrm{eff}}}, \widehat{\Delta\nu}, \widehat{\nu_{\mathrm{max}}}, \hat{R}, \hat{\tau}]$ and $\mathbf{x} = [M, \text{[Fe/H]}, E]$, are outputs from the model at each evolutionary timestamp along the grid. The summary of each variable is shown in Table \ref{table:asfgrid}. Notably, $E$ is a categorical conditional variable and is used to focus on sampling models from a particular evolutionary phase. 
We denote the CNF trained on this grid as $\text{CNF}_{\text{asfgrid}}$.

\section{Results}

\subsection{Emulation Accuracy}

\begin{deluxetable}{c|cc}[t]
\tablecaption{Comparison of the interpolation accuracy of output observables from CNF$_{\mathrm{giant}}$ versus a simple nearest neighbour (NN) approach. Errors are measured as fractional mean absolute residuals. \label{table:flows_vs_nn}}
\tablewidth{0pt} 
\tablehead{ 
\colhead{Observable} & \colhead{CNF$_{\mathrm{giant}}$ (\%)} & \colhead{NN (\%)}
}
\startdata
$T_{\mathrm{eff}}$ & 0.009 & 0.231 \\
$\Delta\nu$ & 0.106 & 1.559 \\ 
$\delta\nu_{01}$ & 3.092 & 14.277 \\
$\delta\nu_{02}$ & 0.339 & 3.387 \\
$\delta\nu_{03}$ & 0.262 & 3.402 \\
$\epsilon$ & 0.388 & 6.278 \\
$R$ & 0.068 & 2.367 \\
$\tau$ & 0.371 & 20.92 \\
\enddata
\end{deluxetable}

We first visualize how well a CNF can emulate the global distribution of model outputs from a grid. Using the original, Sobol-sampled distribution of conditioning variables $\mathbf{x}$ from the MESA grid (defined in Table \ref{table:mesa}), we draw 443,874 samples from $\text{CNF}_{\text{dwarf}}$ and 738,939 samples from $\text{CNF}_{\text{giant}}$. These numbers are chosen to match the number of models in the grid each CNF was trained on. A comparison between the original and emulated distribution of models in the grid is shown in Figure \ref{fig:cornerplot}. Qualitatively, we find that both CNFs are able to closely emulate the distribution of individual output parameters while also capturing global correlations between the parameters despite the high-dimensionality of the grid. 

\begin{figure*}
    \centering
	\includegraphics[width=2.\columnwidth]{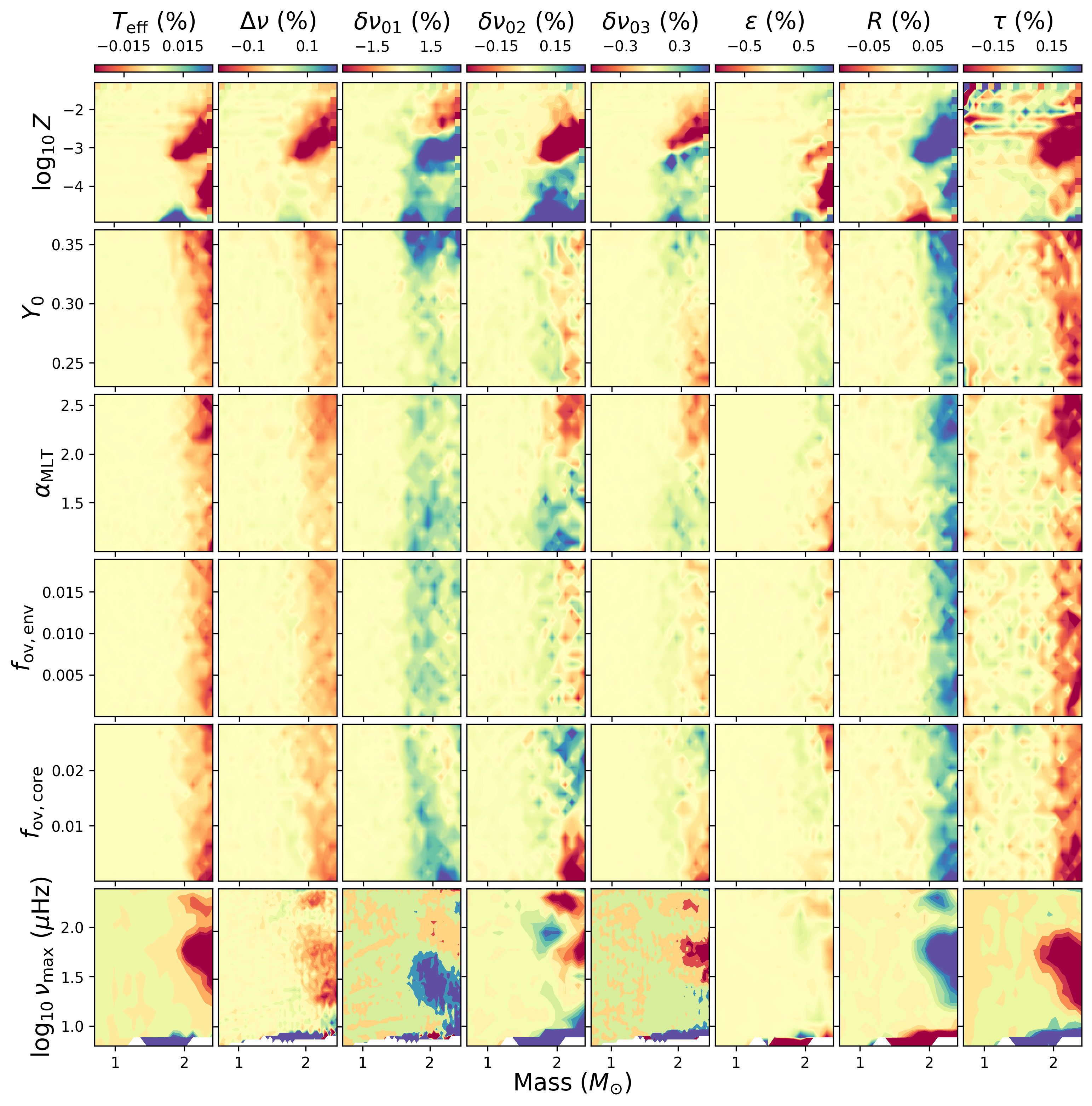}
    \caption{Interpolation errors of output observables from CNF$_{\mathrm{giant}}$ as a function of input parameters. Each panel corresponds to a $20\times20$ grid varied in mass and a second input parameter, which varies from row to row. 
    The color within each panel corresponds to the median fractional deviation (true minus predicted) of a specific output observable at that combination of input parameters, with the specific observable varied across columns. }
    \label{fig:yaguang_model_validation}
\end{figure*}

Because interpolation is a key functionality of the CNF, we investigate the interpolation accuracy CNF$_{\mathrm{giant}}$ and CNF$_{\mathrm{asfgrid}}$, which are the CNFs that are used to infer stellar parameters in Section \ref{section:stellar_parameter_inference}.
To validate the accuracy of CNF$_{\mathrm{giant}}$, we generate a new grid of models with the same input physics and range of input parameters as before, but now using a different random number seed to generate its Sobol sequence. This validation grid also has 8,192 unique combinations of input parameters $\mathbf{x}=$ $[M$, $\log_{10}Z$, $Y$, $\alpha$, $\widehat{f_{\mathrm{ov, env}}}$, $\widehat{f_{\mathrm{ov,core}}}$, $ \widehat{\nu_{\mathrm{max}}}$$]$, but none are identical to the combinations within the grid used to train CNF$_{\mathrm{giant}}$. For each combination of input parameters from the validation grid, we draw 10,000 samples of output observables from CNF$_{\mathrm{giant}}$ and compute each observable's median to obtain a point estimate. We quantify the interpolation accuracy based on how much the point estimates deviate from their corresponding ground truth values from the validation grid. The result of this exercise is shown in Fig. \ref{fig:yaguang_model_validation}, which demonstrates that the interpolation error across most of the grid spans only a fraction of a percent. The interpolations, however, are consistently more erroneous at $M\apprge2.0\,M_{\odot}$ and at $\log_{10}\nu_{\mathrm{max}}\apprle0.5$, which corresponds to models at which their speed of evolution is typically more rapid than the rest of the models in the grid. Therefore, the output distribution of these rapidly-evolving models yield a more `diffuse' density that contribute less to the likelihood optimization and are thus given lower emphasis during training.

A summary of the interpolation accuracy across the whole validation grid for CNF$_{\mathrm{giant}}$ is tabulated in Table \ref{table:flows_vs_nn}. A notable outlier in performance is $\delta\nu_{01}$, which we attribute to the strong degeneracy in such an observable for giant models in across the grid (c.f. Fig. \ref{fig:cornerplot}). To benchmark our results against another method that can perform a multi-dimensional interpolation using multi-dimensional input parameters, we also compare the validation accuracy when using a nearest neighbour interpolation using \texttt{interpolate.griddata} as implemented in \texttt{scipy} \citep{Scipy}. As demonstrated in Table \ref{table:flows_vs_nn}, the errors from CNF are about an order of magnitude smaller across all observables compared to the nearest neighbour method.

\begin{figure}
    \centering
	\includegraphics[width=1.\columnwidth]{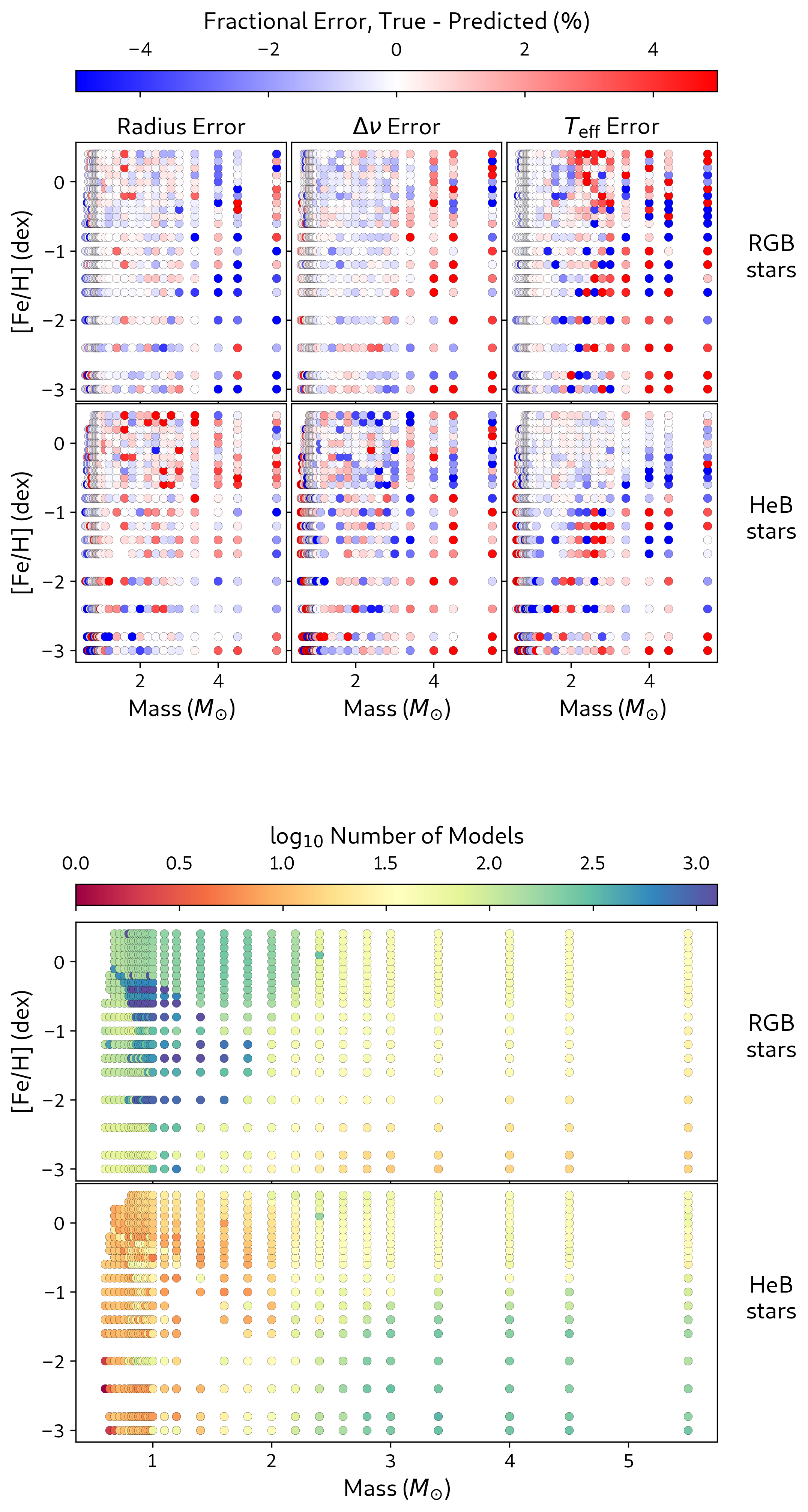}
    \caption{Interpolation errors of output properties [$R$, $\Delta\nu$, $T_{\mathrm{eff}}$] from CNF$_{\mathrm{asfgrid}}$ as a function of input parameters $M$, [Fe/H], and evolutionary state.
    The upper plot corresponds to the 10-fold cross-validation accuracy of CNF$_{\mathrm{asfgrid}}$, while the bottom plot shows the number of models within each evolutionary track of a given $M$ and [Fe/H]. In each plot, models in the hydrogen shell-burning phase (RGB) and in the helium-core burning phase (HeB) of evolution are presented separately.} 
    \label{fig:asfgrid_validation}
\end{figure}

To benchmark CNF$_{\mathrm{asfgrid}}$, we adopt a 10-fold hold out validation strategy in which we train ten independent copies of CNF$_{\mathrm{asfgrid}}$ are trained on 90\% of the tracks in the grid and tested on the remaining 10\%. The train-test split is based on unique values of mass and metallicity, and is split differently for each copy of CNF$_{\mathrm{asfgrid}}$ for this benchmark task. The combination of test results across all ten splits is shown in Fig. \ref{fig:asfgrid_validation}. We find that 67\% of tracks have an absolute fractional interpolation error below 1\%. Meanwhile, the 
the largest interpolation errors often occur within either grid regions that are sparsely populated with models or regions that have large intervals in mass and metallicity between adjacent tracks. These regions generally belong to tracks of high-mass models ($M > 2.5\,M_{\odot}$) or those with low metallicity ([Fe/H] $<$ -2.0 dex). The consistently large errors for helium core-burning stars at the extreme low-mass end ($M=0.6\,M_{\odot}$) of the grid is also a consequence of sparsity, with evolutionary tracks in such regions having 10 models or fewer. Overall, we find that interpolation errors of $R$, $\Delta\nu$,  and $T_{\mathrm{eff}}$ are bounded to a few percent percent within the regions of interest for most of the stars in asteroseismic surveys of the Milky Way, which is $0.7\,M_{\odot}\apprle M\apprle3.0\,M_{\odot}$ and [Fe/H] $>$ -2.5 dex (e.g., \citealt{Pinsonneault_2018, Hon_2021, Theo_2023}).


\subsection{Emulating Evolutionary Tracks}\label{section:conditional_distributions}
To examine the ability of the CNF to interpolate in high dimensions, we use $\text{CNF}_{\text{dwarf}}$ to predict conditional and marginal distributions of evolutionary tracks corresponding in $T_{\mathrm{eff}}-R$ space (a `Hertzsprung-Russell' diagram) as well as the C-D, $\Delta\nu-\epsilon$, and $\delta\nu_{01}-\Delta\nu$ asteroseismic diagrams. These were deliberately selected to highlight the capability of the CNF in emulating varied and complex evolutionary tracks. 

\begin{figure}
    \centering
	\includegraphics[width=1.\columnwidth]{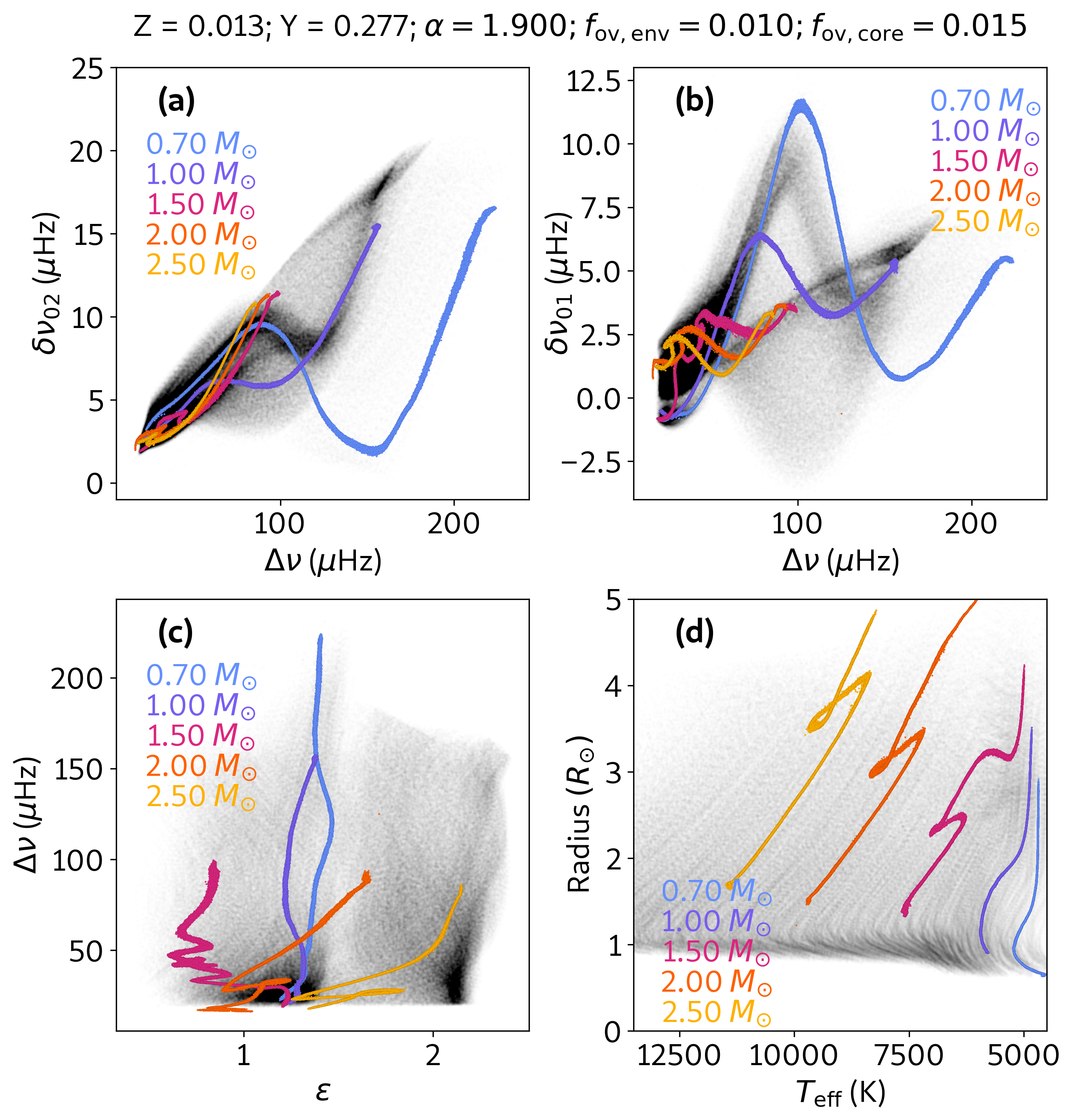}
    \caption{Emulated evolutionary tracks in the grid of MESA dwarf models, predicted as the conditional distribution $p_N(\mathbf{y}| \mathbf{x})$ from CNF$_{\mathrm{dwarf}}$. Here, $\mathbf{x}=(M, Z, Y, \alpha, f_{\mathrm{ov, env}}, f_{\mathrm{ov, core}}$), where $M\in[0.7,1.0,1.5,2.0,2.5]\,M_{\odot}$, with the values of other input parameters listed at the top of the figure. This specific combination of $\mathbf{x}$ is an interpolated value and hence does not exist within the training grid (black). Each panel shows two components from the predicted 9D distribution of output properties in the form of: \textbf{(a)} the asteroseismic C-D diagram, \textbf{(b)} the $\delta\nu_{01}- \Delta\nu)$ diagram, \textbf{(c)}  the $\Delta\nu - \epsilon$ diagram, \textbf{(d)} the `Hertzsprung-Russell' diagram. } 
    \label{fig:varymass}
\end{figure}

The C-D diagram plots the large frequency separation, $\Delta\nu$, against the small frequency separation, $\delta\nu_{02}$ \citep{JCD_1984}. Because $\delta\nu_{02}$ is sensitive to the remaining core hydrogen content of low-mass stars, its value correlates strongly with stellar age when observed as a function of $\Delta\nu$, whose value scales with the root of mean stellar density. As a result, the position of observed stars within the C-D diagram is a powerful diagnostic for the structure and evolution of low-mass stars along the main sequence (e.g., \citealt{Roxburgh_2003, OtiFloranes_2005, White_2011}). The $\delta\nu_{01}-\Delta\nu$ diagram in principle has a similar discriminative power in age to the C-D diagram, albeit with a smaller age sensitivity (e.g., \citealt{Lund_2017}). The phase offset $\epsilon$ carries information related to the structure
of the stellar acoustic mode cavity \citep{Ong_2019} and has been used for its evolutionary diagnostics \citep{JCD_2014} and mode identification properties \citep{White_2012}. Evolutionary tracks in the $\epsilon-\Delta\nu$ diagram, however, are known to vary rapidly and converge for late-stage low-mass main sequence stars, (c.f. \citealt{White_2011, Ong_2019}), which typically limits the applicability of $\epsilon$ to less evolved dwarf stars. Meanwhile, evolutionary tracks of $T_{\mathrm{eff}}$ and $R$ for dwarf stars, similar to their trajectories in the Hertzsprung-Russell diagram, vary smoothly with mass and metallicity but can develop sharp morphological features including the Henyey hook and the dip at the base of the red giant branch \citep{Mengel_1979}. In Figure \ref{fig:varymass}, we present examples of emulated evolutionary tracks in these four diagrams. For each specific combination of $\mathbf{x}$, we draw a total of $10^6$ samples from CNF$_{\mathrm{dwarf}}$ to yield the conditional distribution $p_N(\textbf{y}|\textbf{x})$. In general, we find that CNF$_{\mathrm{dwarf}}$ is capable of faithfully emulating complex patterns in evolutionary tracks, including intersecting and looping trajectories, sharp transitions, and oscillatory variations. 

\begin{figure*}
    \centering
	\includegraphics[width=2.\columnwidth]{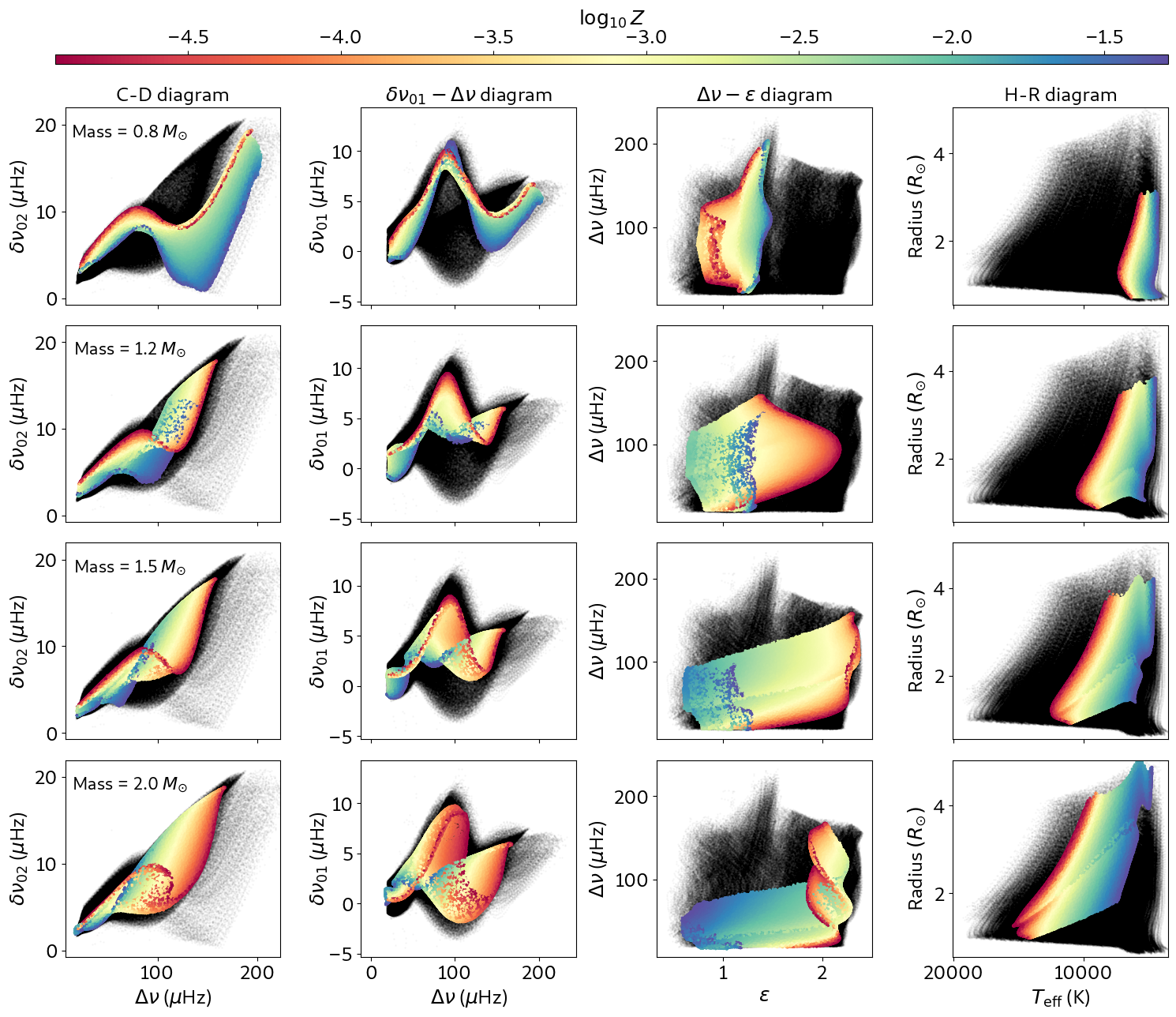}
    \caption{Continua of emulated evolutionary tracks over metallicity by CNF$_{\mathrm{dwarf}}$, shown as the distribution of $\mathbf{y}$ given $\mathbf{\Bar{x}} = (M, Y, \alpha, f_{\mathrm{ov, env}}, f_{\mathrm{ov, core}})$ marginalized over $Z$, or $p_N(\mathbf{y}|\mathbf{\Bar{x}})\approx\int p(\mathbf{y}| Z, \mathbf{\Bar{x}})\, p(Z)\,dZ$. Each row shows CNF$_{\mathrm{dwarf}}$ conditioned on a specific value of stellar mass, with the adopted values of the other conditional values in $\mathbf{\Bar{x}}$ the same as those used in Figure \ref{fig:varymass}. Each column represents a different type of diagram for which evolutionary tracks are commonly visualized. Black points in the background correspond to models across the whole grid, which illustrates the region of parameter space potentially covered by the flow's predictions.}
    \label{fig:dwarf_feh_dependence}
\end{figure*}

\begin{figure*}
    \centering
	\includegraphics[width=2.\columnwidth]{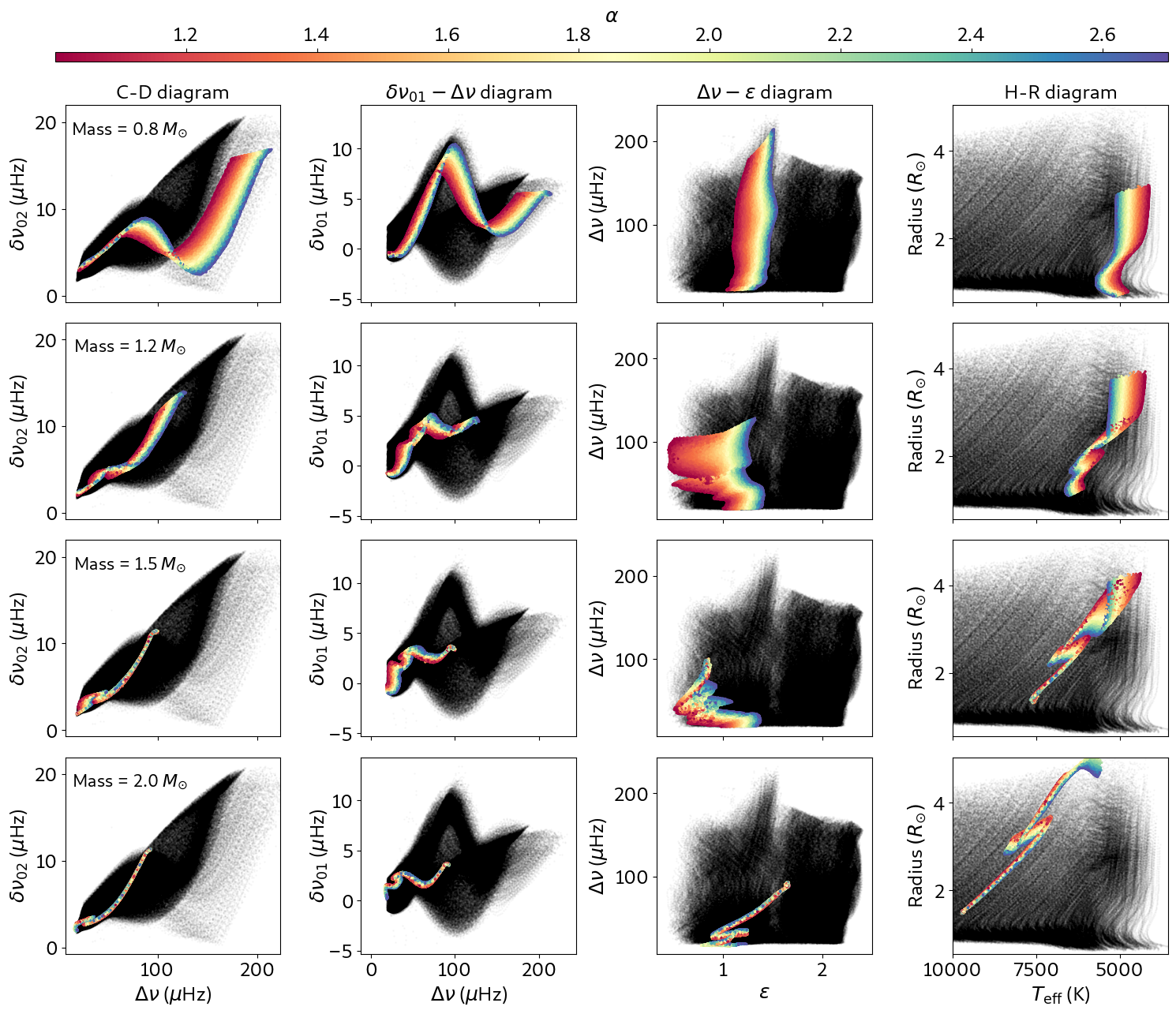}
    \caption{Same as Figure \ref{fig:dwarf_feh_dependence}, except that the marginalization is now performed over the mixing length parameter $\alpha$. Therefore, the distributions predicted from CNF$_{\mathrm{dwarf}}$ here indicate $p_N(\mathbf{y}|\mathbf{\Bar{x}})\approx\int p(\mathbf{y}| \alpha, \mathbf{\Bar{x}})\, p(\alpha)\,d\alpha$, with $\mathbf{\Bar{x}} =(M, Z, Y, f_{\mathrm{ov, env}}, f_{\mathrm{ov, core}})$ and $Z=0.013$.}
    \label{fig:dwarf_alpha_dependence}
\end{figure*}

\begin{figure}
    \centering
	\includegraphics[width=1.\columnwidth]{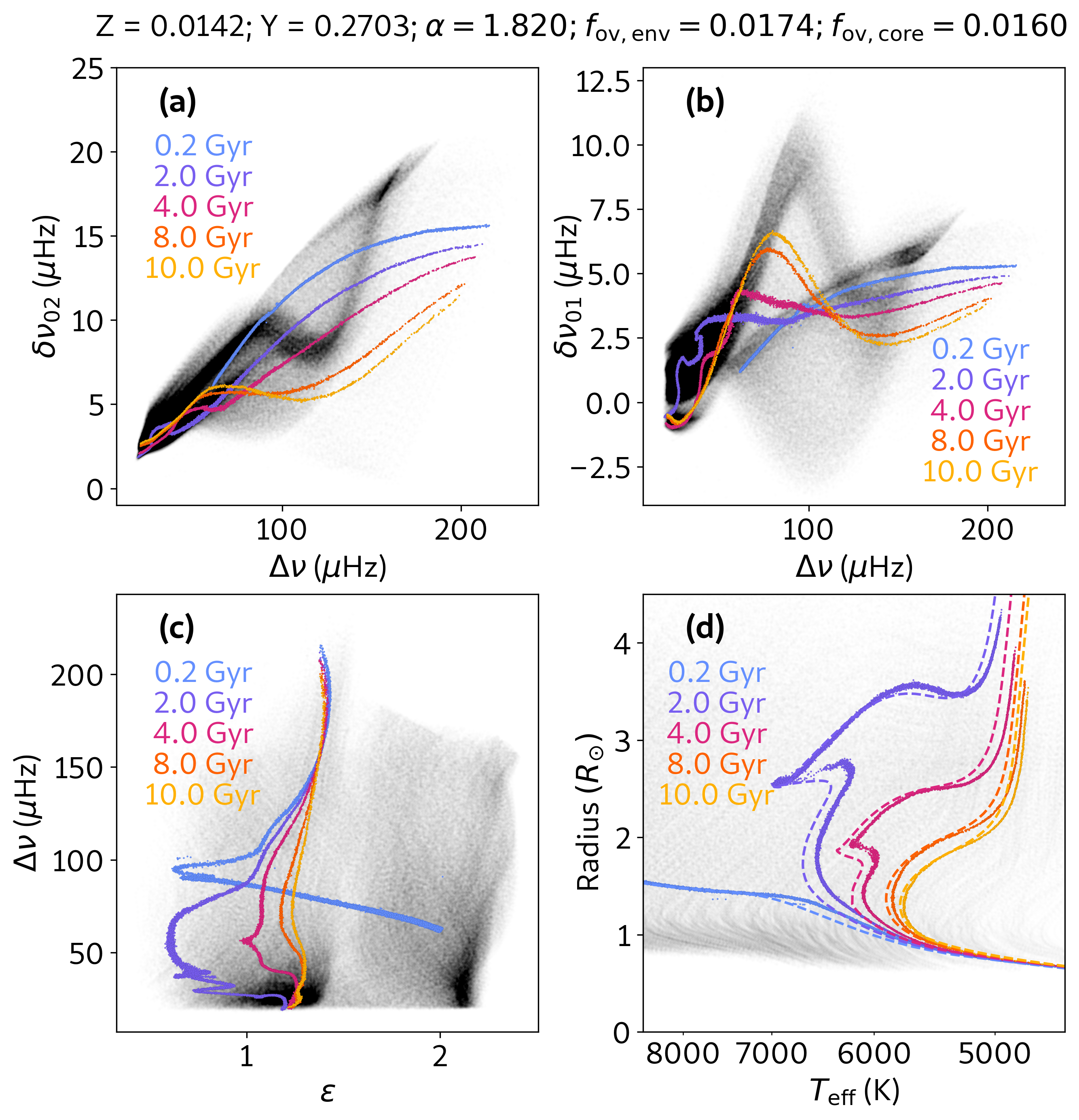}
    \caption{Same as Fig. \ref{fig:varymass}, except here CNF$_{\mathrm{dwarf}}$ predicts isochrones instead of evolutionary tracks, such that $p_N(\mathbf{\Bar{y}}| \tau, \mathbf{x})$. By marginalizing over mass such that $M\sim\mathcal{U}(0.7, 2.5) M_{\odot}$, we set $\tau\in[0.2, 2, 4, 8, 10]\,$Gyr to be a conditional variable, where the vector $\mathbf{\Bar{y}}$ is $\mathbf{y}$ excluding $\tau$. Other components of $\mathbf{x}$ are listed above the figure.
    In practice, $p_N(\mathbf{\Bar{y}}| \tau, \mathbf{x})$ is obtained by drawing samples from $p_N(\mathbf{y}|\mathbf{x})$ and accepting samples with an age within a user-defined tolerance (here 0.01 Gyr) of $\tau$. The dashed lines in panel (d) correspond to MIST isochrones \citep{Dotter_2016}, which have the same input parameters $\mathbf{x}$ but adopt a different set of input physics relative to the MESA grid in this study.}
    \label{fig:varyage}
\end{figure}

\begin{figure}
\begin{interactive}{animation}{video file name}
	\includegraphics[width=1.\columnwidth]{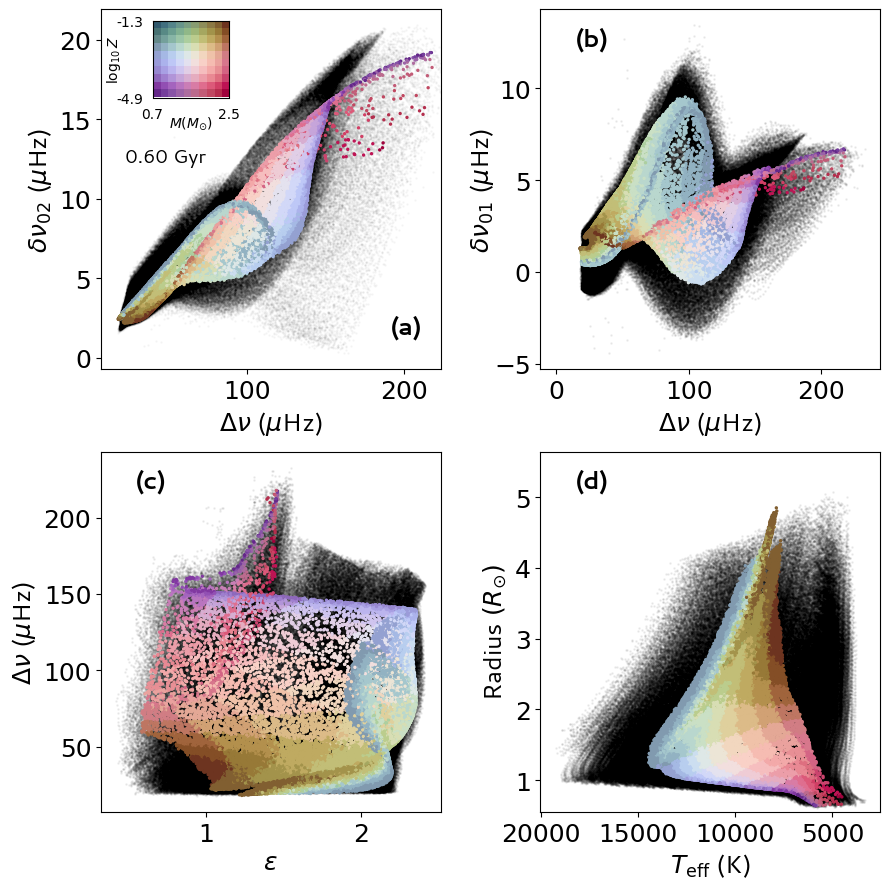}
 \end{interactive}
\caption{Continua of isochrones in the form of emulated distributions from CNF$_{\mathrm{dwarf}}$ that are conditioned on age and marginalized over both metallicity ($Z$) and mass ($M$). The value of age ($\tau$) is varied first from $0.05-1\,$Gyr in increments of $0.05\,$Gyr, then from $1-15\,$Gyr in increments of $0.15\,$Gyr. 
The distribution at each age timestamp represents 
$p_N(\mathbf{\Bar{y}}|\mathbf{\Bar{x}}, \tau)\approx  \int\int p(\mathbf{\Bar{y}}| Z, M, \tau, \mathbf{\Bar{x}})\, p(Z)\,p(M|\tau)\,dZ\,dM$, where $\mathbf{\Bar{y}}$ is $\mathbf{y}$ excluding $\tau$. Meanwhile, the vector $\mathbf{\Bar{x}}$ is $\mathbf{x}$ excluding $M$ and $Z$, with other input parameter values are the same as those in Figure \ref{fig:varymass}. An animated version of this figure is available in the HTML version of the final article.}
\label{fig:animation}
\end{figure}

\begin{figure*}
    \centering
	\includegraphics[width=2.\columnwidth]{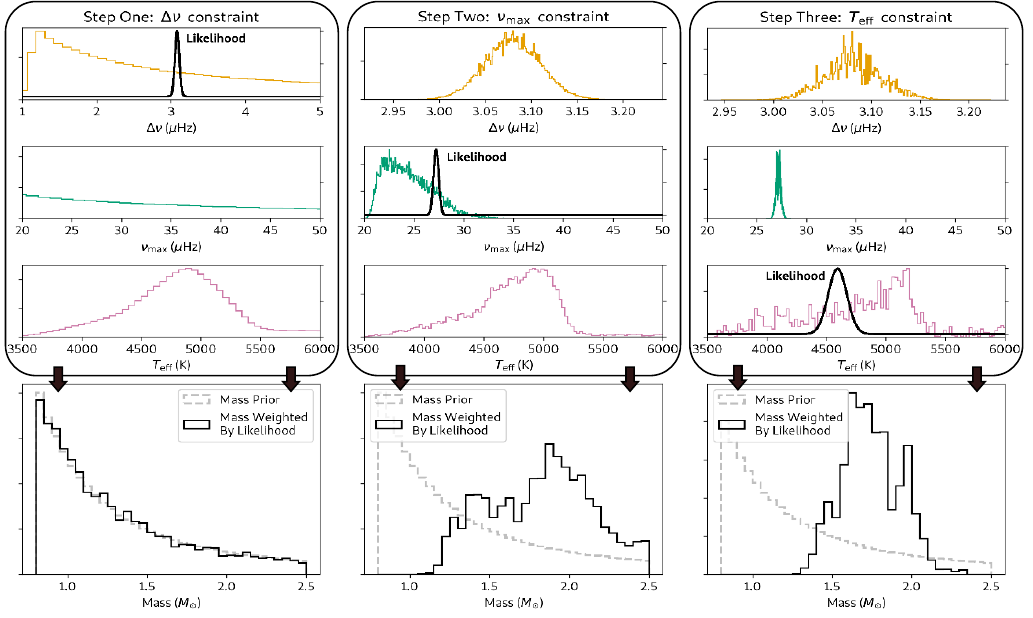}
    \caption{Sampling a mass posterior distribution from CNF$_{\mathrm{asfgrid}}$ using constraints from observables $\Delta\nu$, $\nu_{\mathrm{max}}$, and $T_{\mathrm{eff}}$. The upper three panels in each column (steps one to three) contain histograms of emulated observables from the CNF using a mass prior that is shown in the bottom panels. In step one, a constraint of $\Delta\nu=3.08\pm0.03\,\mu$Hz forms the applied likelihood. In step two, a constraint of $27.2\pm0.3\,\mu$Hz is applied. In step three, a constraint of $4593\pm78\,$K is applied. The sequential applications of constraints are for illustrative purposes only; the ordering of constraints from observables are arbitrary in practice.} In the bottom panel of each step, samples from the prior mass distribution are directly weighted by the applied likelihood and are resampled using SIR to approximate a posterior distribution. Note that from step two onwards, the distributions of emulated observables also change due the weighting of a likelihood from the previous step.
    \label{fig:sir_stages}
\end{figure*}

\subsection{Emulating a Continuum of Tracks}\label{section:marginal_distributions}
Importantly, sampled evolutionary tracks from the CNFs smoothly vary across variations in input grid parameters, in alignment with expectations of stellar evolution. The smoothness of the CNF emulation is attributed to the inherent smoothness of the CNF's latent space and the ability of its neural network to interpolate across the space of input parameters. We visualize the smoothness of the emulation by generating a continuum of evolutionary tracks, which is done by marginalizing over the range of specific input parameters.

We present a continuum of evolutionary tracks varied in metallicity in Fig. \ref{fig:dwarf_feh_dependence}. Metallicity affects radiative opacities, equations of state, and nuclear reaction rates \citep{Mowlavi_1998}, resulting a substantial influence on the observable properties of stars as they evolve. As demonstrated in Fig. \ref{fig:dwarf_feh_dependence}, this influence across a continuous range of observables can be distinctly visualized using the CNF by smoothly interpolating across $Z$. 
This marginalization over $Z$ requires minimal modification from the conditional distribution scenario (Fig. \ref{fig:varymass}) by simply sampling with $\log_{10}Z\sim\mathcal{U}(-4.934, -1.291)$, rather than fixing it to a specific value.   

To present another example, we next display continuous variations from varying the mixing length parameter, $\alpha$, in Figure \ref{fig:dwarf_alpha_dependence}. 
The dimensionless quantity $\alpha$ defines a characteristic distance that a parcel of convective material can travel adiabatically within the star before blending with its surrounding medium. Hence, it is particularly important for characterizing the efficiency of energy transport in stars with superadiabatic convective zones, which includes stars with $M\apprle1.2M_{\odot}$ and giant stars \citep{Bohm_1958, Henyey_1965}.
The impact of varying $\alpha$ on stellar observables is of particular interest, especially given mounting evidence the value of $\alpha$ requires calibration across stars (e.g., \citealt{Joyce_2018, Joyce_2018b, Li_2018}) and contains dependencies on stellar composition and evolution (e.g., \citealt{Tanner_2013, Tayar_2017})
In Fig. \ref{fig:dwarf_alpha_dependence}, we observe the expected global systematic shift of evolutionary tracks to cooler temperatures in the H-R diagram with lower $\alpha$ due to a lower energy transport efficiency. This effect is markedly less pronounced on the main sequence for stars with $M>1.2M_{\odot}$ as a consequence of their interior structure comprising convective cores with radiative envelopes. As described by \citet{Joyce_2023} and observed in the Figure, the $T_{\mathrm{eff}}$ of these stars are nonetheless modified with varying $\alpha$ once they evolve past the main sequence turn-off, corresponding to the development of a convective envelope during the subgiant phase and beyond. Overall, 
Figs. \ref{fig:dwarf_feh_dependence} and \ref{fig:dwarf_alpha_dependence} demonstrate the ability of CNF-based emulation to distinctly quantify the extent of change of evolutionary tracks with respect to specific input parameters. This ability allows disentangling variations in observables as a result of a change in only one specific parameter, which can be insightful in guiding the optimization of model fitting approaches.


\subsection{Emulating Isochrones}\label{section:output_constraints}
The CNF can additionally accept constraints on output properties to emulate conditional distributions, which is performed by drawing samples from the CNF and retaining only those that have emulated properties that match user-specified constraints. Since we have stellar age as an output property ($\tau\in\mathbf{y}$) across our trained CNFs, we can emulate isochrones by imposing an age restriction upon generated samples from $p_N(\mathbf{y}|\mathbf{x})$. Fig. \ref{fig:varyage}a-c demonstrates that emulated isochrones with CNF$_{\mathrm{dwarf}}$ can indeed emulate mono-age samples of models within the grid, evidenced by the observed distributions in the radius-temperature diagram and the stratification of tracks in $\delta\nu_{01}$ and $\delta\nu_{02}$ relative to $\Delta\nu$, which is the defining property of the C-D diagram for age-dating low-mass dwarf stars (e.g., Fig. 7 of \citealt{White_2011}, Fig. 4 of \citealt{Bellinger_2019}). Fig. \ref{fig:varyage}d compares the emulation of the CNF with MIST isochrones \citep{Dotter_2016}, showing that the emulation qualitatively behaves like standard isochrones. The offsets between the MIST isochrones with the CNF emulation are expected, given differences in the adopted prescriptions and microphysics between the MIST grid \citep{Choi_2016} and ours.

The ability to emulate the stellar properties of mono-age models is valuable for studying the properties of co-eval systems such as binary stars and stellar clusters, given that fitting the observables of such systems commonly involve varying isochrones across a range of input parameters like mass and metallicity (e.g., \citealt{An_2007}). Similar to our approach with evolutionary tracks, it is straightforward to emulate a continuum of isochrones spanning a range of input parameters by marginalizing over components in $\mathbf{x}$ at specific timestamps. We show such an emulation that marginalizes over both mass and metallicity in Fig. \ref{fig:animation} across the age range of $0.1-15\,$Gyr, which highlights the CNF's ability to capture continuously evolving variations in isochrones with age.

Overall, the CNF emulator shows great potential as an instructive tool for visualizing how input parameters and output properties within a grid of stellar models interact with one another. To highlight this aspect, we present a publicly accessible interactive demonstration of CNF$_{\mathrm{dwarf}}$ that allows users to explore the conditional distributions in the evolutionary diagrams presented in this work, with details in Appendix \ref{appendix:interactive}.

\subsection{Bayesian Inference by Sampling/Importance Resampling}\label{section:stellar_parameter_inference}
Because the CNF interprets the distribution of models within a grid as a density, it can act as a generative prior in Bayesian inference tasks. Given samples $\mathbf{x}$ drawn from a prior distribution $ p(\mathbf{x})$, the role of the CNF is to directly map the samples into $\mathbf{y}$, which exist in the space of output properties (or observables). These samples can subsequently be weighted using some likelihood of observed data, $p(\mathbf{y}|\mathbf{x})$. Because the CNF provides a one-to-one mapping between samples $\mathbf{x}$ and samples $\mathbf{y}$, the weights from the likelihood can be directly applied to samples drawn from the prior distribution. Following Bayes theorem, we would expect that weighting the prior with the data likelihood following this approach would thus provide an approximation to a posterior distribution. 

Formally, this approach is known as Sampling/Importance Resampling (SIR, \citealt{Rubin_1988}). As an approach to Bayesian inference, we first assume there exists a positive function $k(\phi)$ that can be expressed following Bayes' theorem as $k(\phi)$ = $\mathcal{L}(\phi)\,p(\phi)$, where $p(\phi)$ is a prior density and $\mathcal{L}(\phi)$ is the likelihood function given by observed data. Following the description by \citet{Smith_1992}, 
SIR seeks to estimate the target probability density $h(\phi) = k(\phi)/\int\,k(\phi)\,d\phi$ by the following:
\begin{enumerate}
    \item Draw samples $\phi_i,$ with $i=1,2,\cdots,n$ from a proposal density $g(\phi)$.
    \item Calculate sample importance weights $q_i$ = $\omega_i/\sum_{j=1}^n\omega_j$, where $\omega_i = k(\phi_i)/g(\phi_i)$.
    \item Draw $N$ samples $\phi^*$ from the discrete distribution over $\{\phi_1, \phi_2, \cdots, \phi_n\}$ with probability mass $q_i$ on $\phi_i$.
\end{enumerate}
The drawn samples $\phi^* = \{\phi^*_1, \phi^*_2, \cdots, \phi^*_N\}$ will consequently be approximately distributed following the target density $h(\phi)$. The selection of the proposal density $g(\phi)$ follows from standard importance sampling, in that we typically seek a $g(\phi)$ that covers the support of $h(\phi)$. 

Because $k(\phi_i)=\omega_i\,g(\phi_i)$ and $k(\phi)$ = $\mathcal{L}(\phi)\,p(\phi)$, we may equate the prior density $p(\phi)$ with the proposal density $g(\phi)$ to obtain a special case for which the importance weights $q_i$ form the data likelihood: $q_i = \mathcal{L}(\phi_i)/\sum_{j=1}^n\mathcal{L}(\phi_j)$. In other words, the posterior density can be approximated by simply weighting samples drawn from the prior by a likelihood given by the observed data. 

In summary, using SIR, we can approximate Bayesian posterior densities of input parameters $\mathbf{x}$ drawn from a prior density.  The advantages of using the CNF for Bayesian inference are the following:
\begin{itemize}
    \item \textbf{The resolution of the posterior distribution is not limited by the inherent resolution of the original grid}. 
    Techniques for computing posterior distributions from a grid of models commonly involve sampling or integrating over a volume of the input parameter space discretized by the grid itself, which can heavily depend on the concentration of models within that volume. In contrast, Sections \ref{section:conditional_distributions} and \ref{section:marginal_distributions}, demonstrate that the CNF can smoothly interpolate across $\mathbf{x}$, such that a prior distribution $p(\textbf{x})$ passed through the CNF yields a continuous space of emulated observables that is not limited by the original grid's discretization.
    \item \textbf{There exists a one-to-one mapping between samples from the prior distribution, samples in the observable space, and samples from the posterior distribution}. This provides a direct approach for examining how specific observed data points, including outliers or extreme values, shape the resulting posterior distribution. An example of this mapping is shown in Fig. \ref{fig:sir_stages}, in which weighting from a data likelihood is applied sequentually to samples drawn from a prior distribution.
\end{itemize}

In this study, we perform stellar parameter inference using CNF-emulated grids on two tasks. The first uses CNF$_{\mathrm{giant}}$ to estimate the ages and input parameters of \textit{Kepler} red giants in the open clusters NGC 6791 and 6891. The second task uses CNF$_{\mathrm{asfgrid}}$ to re-estimate the masses and radii of \textit{Kepler} red giants analyzed by \citet{Yu_2018}. In each task, we assume that the observables $y_{\mathrm{obs}}$ forming the vector $\mathbf{\Bar{y}}$ are mutually independent and Gaussian distributed, such that the likelihood $\mathcal{L}$ can be expressed by
\begin{equation}\label{eqn:likelihood}
    \mathcal{L} = \prod_{y\,\in\, \mathbf{\Bar{y}}}\frac{1}{\sigma_{y_{\mathrm{obs}}} \sqrt{2\pi}}\,\text{exp}\frac{(y_{\mathrm{obs}} - y_{\mathrm{CNF}})^2}{2\sigma_{y_{\mathrm{obs}}}^2}
    ,
\end{equation}
\noindent where $y_{\mathrm{CNF}}$ are the predicted observables from the CNF and $y_{\mathrm{obs}}$ are the uncertainties of the observables.

Alongside the trained CNFs presented in this study, we present the \texttt{modelflows}\footnote{\url{https://github.com/mtyhon/modelflows}} Python package \citep{modelflows} that implements SIR for inferring stellar parameters.

\begin{figure*}
    \centering
	\includegraphics[width=2.\columnwidth]{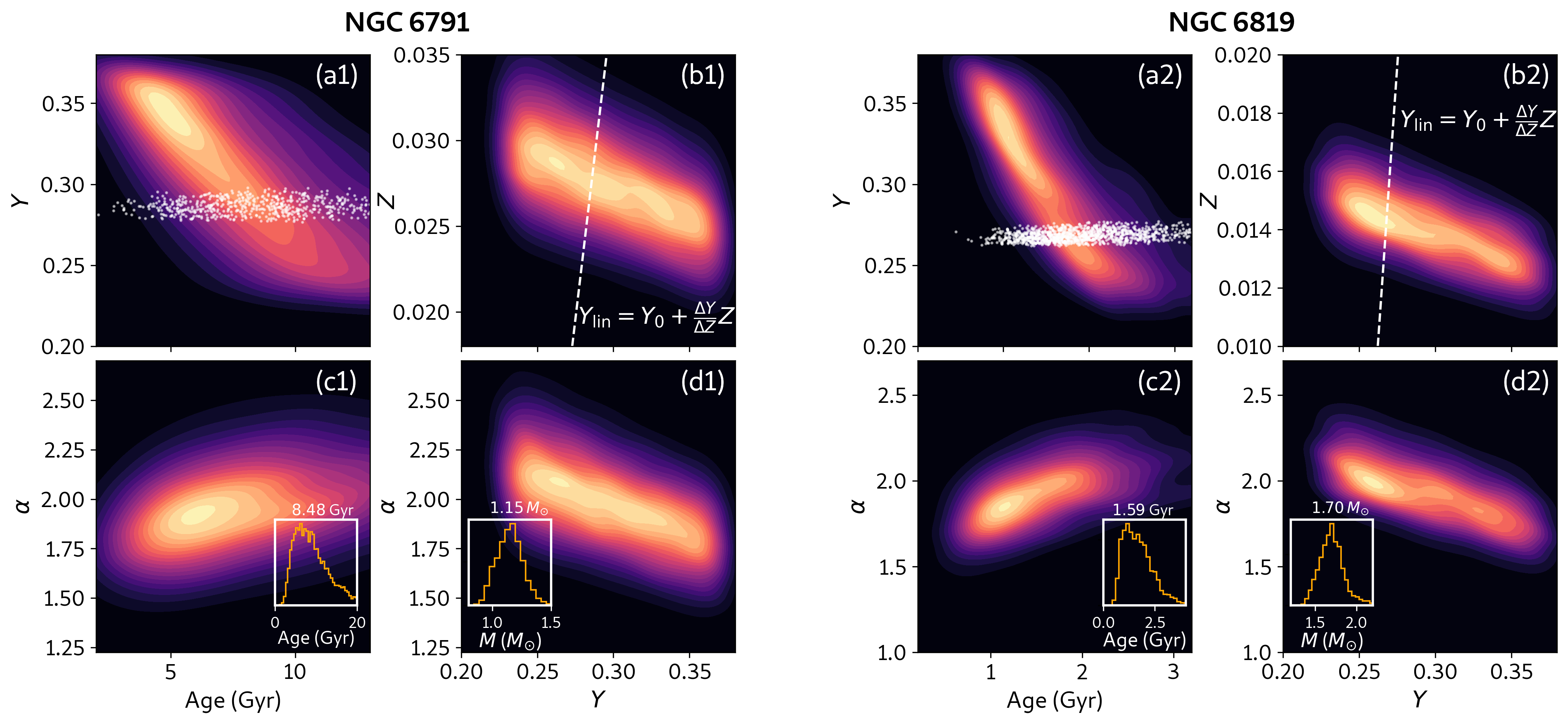}
    \caption{2D kernel density estimates of stacked posterior distributions of input parameters for red giant branch stars in open clusters NGC 6791 and NGC 6819, as inferred by CNF$_{\mathrm{giant}}$. Insets in panels (c1) and (c2) correspond to the posterior distribution of stellar age for NGC 6791 and NGC 6819, respectively, while insets in panels (d1) and (d2) are posterior distributions for mass. The value above each inset panel corresponds to the median of the plotted distribution. The dashed lines in panels (b1) and (b2) shows the region in the $Z-Y$ parameter space explored using a linear helium enrichment relation $Y_{\mathrm{lin}}$, here shown with $Y_0 = 0.249$ and $\Delta Y/\Delta Z = 1.33$. White points in (a1) are samples from the joint posterior distribution for NGC 6791 that have $| Y-Y_{\mathrm{lin}} < 0.001|$, which demonstrates the age boundaries implicitly assumed when adopting $Y_{\mathrm{lin}}$. White samples in (b2) are similar, but for NGC 6819.}
    \label{fig:openclusters}
\end{figure*}

\subsection{Stellar Parameter Inference: Kepler Open Clusters} \label{section:open_clusters}
Here, we aim to demonstrate the uncertainties encountered in calculating stellar ages as a consequence of the competing influences across the grid's input parameters on the ages of models. We perform this demonstration by applying CNF$_{\mathrm{giant}}$ on red giant branch stars within the open clusters NGC 6791 and NGC 6819 observed by the \textit{Kepler} mission. We use broad, uniformly sampled priors for $Z$, $Y$, $\alpha$, $f_{\mathrm{ov, env}}$, and $f_{\mathrm{ov, core}}$, while the distribution for $M$ was drawn from a standard Salpeter initial mass function independent of age and metallicity. 
Given the lack of mass loss in our grid of models, the present-day prior distribution of masses is assumed to follow the initial mass prior distribution.
We obtain posterior distributions for each star by sampling from the prior distributions, emulating the grid properties using CNF$_{\mathrm{giant}}$, and applying SIR. Using this approach, samples are drawn until the discrete distribution $\phi^*$ over $\{\phi_1, \phi_2, \cdots, \phi_n\}$ contains a minimum of 50,000 non-zero importance weights. Because the CNF maintains a one-to-one mapping between input parameters and output observables (as described in Section \ref{section:stellar_parameter_inference}), we obtain posterior distributions of $\tau$ simply by examining the ages of samples forming the posterior distributions for $\Bar{\mathbf{x}}$.   

For NGC 6791, we adopt observables $\Delta\nu$, $\nu_{\mathrm{max}}$, and $T_{\mathrm{eff}}$ of the 27 red giant branch stars from the analysis by \citet{Mckeever_2019} for our likelihood function $\mathcal{L}$. We additionally impose a metallicity constraint for the cluster of [Fe/H] = $+0.35\pm0.04$ dex based on the Open Cluster Chemical Abundances and Mapping Survey (OCCAM, \citealt{Donor_2020}), which presented abundance analyses from APOGEE Data Release 16. For NGC 6819, we use measurements of $\Delta\nu$, $\nu_{\mathrm{max}}$, $\delta\nu_{02}$ and $T_{\mathrm{eff}}$ for 20 red giant branch stars from the study by \citet{Handberg_2017}, removing any reported overmassive stars or non-cluster members in the process. We adopt a bulk metallicity constraint of [Fe/H] = $+0.05\pm0.03$ dex from OCCAM.

\begin{figure*}
    \centering
	\includegraphics[width=2.\columnwidth]{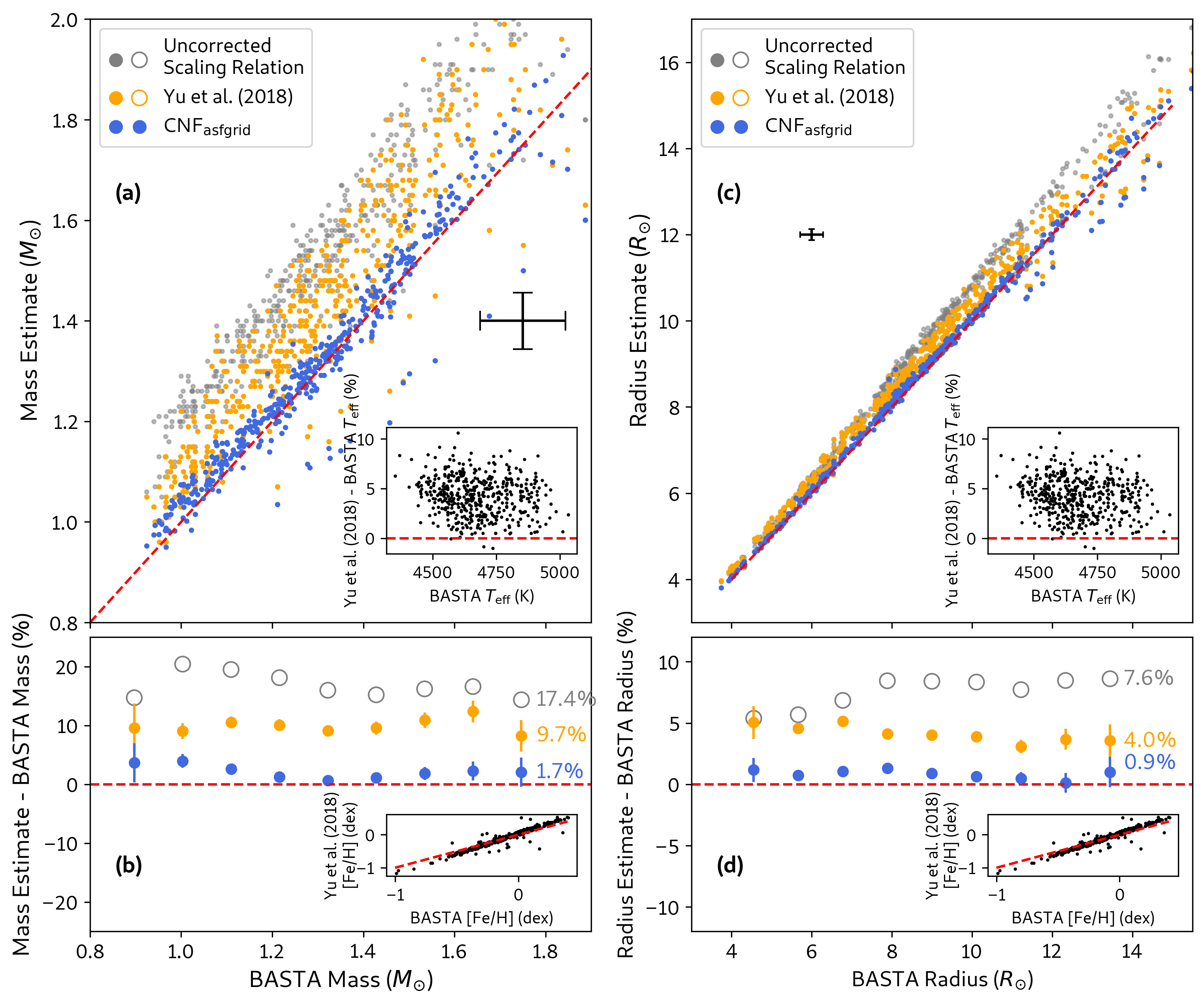}
    \caption{Mass (a-b) and radius (c-d) estimates from CNF$_{\mathrm{asfgrid}}$ for 1,199 \textit{Kepler} field red giant branch stars. The one-to-one relation (red dotted line) is shown with respect to results from the BAyesian STellar Algorithm (BASTA) reported by \citet{Silva-Aguirre_2018}. Also shown are mass and radius estimates using the uncorrected scaling relations in gray (Eqn. \ref{eqn:scaling_relations}) as well as the values reported from the \citet{Yu_2018} catalog. Fractional residual plots are shown in panels (b) and (d), with the median fractional deviation across each estimate source labelled accordingly. Representative errorbars are shown in panels (a) and (c), with inset panels in (a-d) showing differences in spectroscopic parameters $T_{\mathrm{eff}}$ and [Fe/H] adopted by the \citet{Silva-Aguirre_2018} and \citet{Yu_2018} studies.}
    \label{fig:asfgrid_basta}
\end{figure*}

By combining samples from the posterior distributions across stars from each open cluster, we present the heatmaps describing correlations between $Z$, $Y$, $\alpha$, and age in Fig. \ref{fig:openclusters}. We observe the following:

\begin{itemize}
    \item The strong influence of initial helium abundances $Y$ on stellar ages $\tau$. An increase in $Y$ results in an increase to a star's mean molecular weight, which increases its luminosity and decreases its lifetime along the main sequence. Panels (a1) and (a2) in Fig. \ref{fig:openclusters} show that in the absence of a strong prior in $Y$, there exists models with a diverse range of initial helium abundances whose output properties may provide reasonable fits to the seismic and spectroscopic parameters of the observed stars. The combined spread in ages for these models spans a range of about 10 Gyr for NGC 6791 and about 2.5 Gyr for NGC 6819 within our grid of models. 
    \item Despite such large variations in age, the combined mass posterior distributions are reasonably well-constrained, with a spread of about $0.12\,M_{\odot}$ for both clusters. The median mass of $M=1.15\,M_{\odot}$ for NGC 6791 here is in good agreement with the reported $M=1.15\pm0.01\,M_{\odot}$ from \citet{Mckeever_2019}.
    Meanwhile, the $M=1.70\,M_{\odot}$ median mass reported for NGC 6819 here is skewed to a higher mass (and thus younger age) relative to its value of $M=1.61\pm0.02\,M_{\odot}$ reported from literature \citep{Miglio_2012,Handberg_2017}. This difference can be attributed to the additional degrees of freedom in our grid for examining models at the extreme ends of $Y$ and $\alpha$. 
    \item An expected anti-correlation between $Z$ and $Y$ in panels (b1) and (b2). Given that the ratio $Z/X$ is constrained by the observed bulk metallicity of each cluster, an increase in $Y$ must lead to diminishing $Z$, with its spread determined by the uncertainty in [Fe/H]. In these panels, we also visualize the linear helium enrichment relation, a commonly used prior for initial helium abundances, and how such a prior reduces the parameter space for stellar ages in panels (a1) and (a2). 
    \item A positive correlation of mixing length $\alpha$ with stellar age in panels (c1) and (c2). This is a trend also observed by other model-fitting studies (e.g., \citealt{Joyce_2018b}, Li et al. 2024 in review), which emerges as a result of the competing influence between $Y$ and $\alpha$ in the fitting of stellar observables. For instance, the increase in $T_{\mathrm{eff}}$ due to an increase in $\alpha$ in a model (c.f. Fig. \ref{fig:dwarf_alpha_dependence}) needs to be balanced with the decrease in $T_{\mathrm{eff}}$ from a higher $Y$, among other observables. This anti-correlation between $Y$ and $\alpha$ \textemdash{} evident in panels (d1) and (d2) \textemdash{} indicates that assuming a constant value of $\alpha$ indirectly sets a prior on $Y$, and this indirectly influences the age range of models that may be explored from the grid. 
\end{itemize}

To date, there exists a broad range of ages that have been reported for NGC 6791 ($6-13$ Gyr) and for NGC 6819 ($1.5-3$ Gyr), which partially stems from the wide variety of input parameters priors for grid-based modelling (see e.g., discussions and comparisons by \citealt{Basu_2011, Mckeever_2019, Li_2023}). Fig. \ref{fig:openclusters} shows that without enforcing a strong prior across input parameters, the age of a co-eval population determined across its individual members can be highly ambiguous. Rather, a more judicious approach may be to explicitly include a goodness-of-fit metric that penalizes strong differences in input parameters across co-eval members, such as those implemented by \citet{Joyce_2018b}. These results are also indicative of the insufficiency of spectroscopic and global asteroseismic observables alone in placing strong constraints on $Y$, $\alpha$, and stellar ages. Rather, further independent constraints are needed from cluster-bound eclipsing binaries \citep{Brogaard_2012, Mckeever_2019, Brogaard_2021}, and also from the measurement of helium abundances using detailed asteroseismic measurements of interior stellar structure \citep{Verma_2019, Nsamba_2021, Verma_2022}. \\

\begin{deluxetable}{c|ccc}[t]
\tablecaption{Estimates of 15,388 \textit{Kepler} red giants from the \citet{Yu_2018} catalog based on the emulation of the grid accompanying AsfGrid. Uncertainties are reported as the deviation of the 16th and 84th percentile values from the median. The column `Ev' refers to the adopted evolutionary stage of the star, where 1 is for hydrogen shell-burning (RGB) red giants, whereas 2 is for helium core-burning (HeB) stars. The full version of this table is available in a machine-readable format in the online journal, with a portion shown here for guidance regarding its form and content. \label{table:kepler_rg}}
\tablewidth{0pt} 
\tablehead{ 
\colhead{KIC} & \colhead{ Mass ($M_{\odot}$)} & \colhead{Radius ($R_{\odot}$)} & \colhead{Ev}
}
\startdata
757137 & $1.417^{+0.123}_{-0.089}$ & $12.799^{+0.486}_{-0.247}$ & 1 \\
892760 & $0.959^{+0.066}_{-0.067}$ & $10.459^{+0.338}_{-0.308}$ & 2\\
893214 & $1.500^{+0.075}_{-0.070}$ & $11.185^{+0.221}_{-0.216}$ & 1\\
1026084 & $1.533^{+0.167}_{-0.105}$ & $11.204^{+0.481}_{-0.263}$ & 2\\
$\cdots$ & $\cdots$ & $\cdots$ & $\cdots$ \\
12885373 & $1.264^{+0.044}_{-0.084}$ & $9.116^{+0.163}_{-0.176}$ & 1\\
12934574 & $1.020^{+0.063}_{-0.065}$ & $10.743^{+0.264}_{-0.261}$ & 2\\
\enddata
\end{deluxetable}

\begin{figure*}
    \centering
	\includegraphics[width=2.\columnwidth]{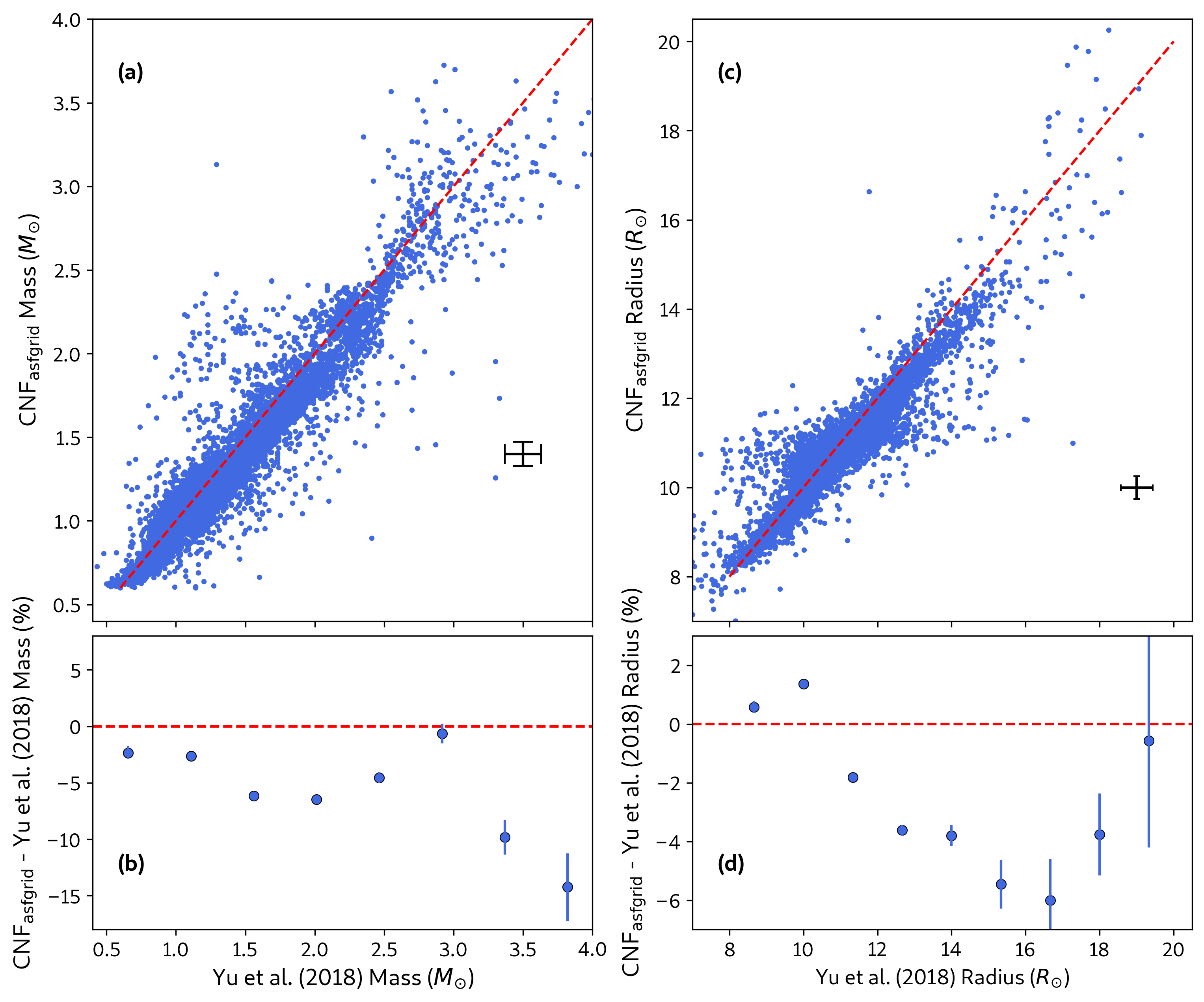}
    \caption{Mass (a-b) and radius (c-d) estimates from CNF$_{\mathrm{asfgrid}}$ for 7,703 helium core-burning (CHeB) \textit{Kepler} field red giant stars in comparison to estimates reported by \citet{Yu_2018}. The one-to-one relation is shown with the red dotted line. Both results use identical asteroseismic and spectroscopic input data, differing only in their methods to retrieve values from the same grid of models.}
    \label{fig:asfgrid_clump_comparison}
\end{figure*}

\begin{figure}
    \centering
	\includegraphics[width=1.\columnwidth]{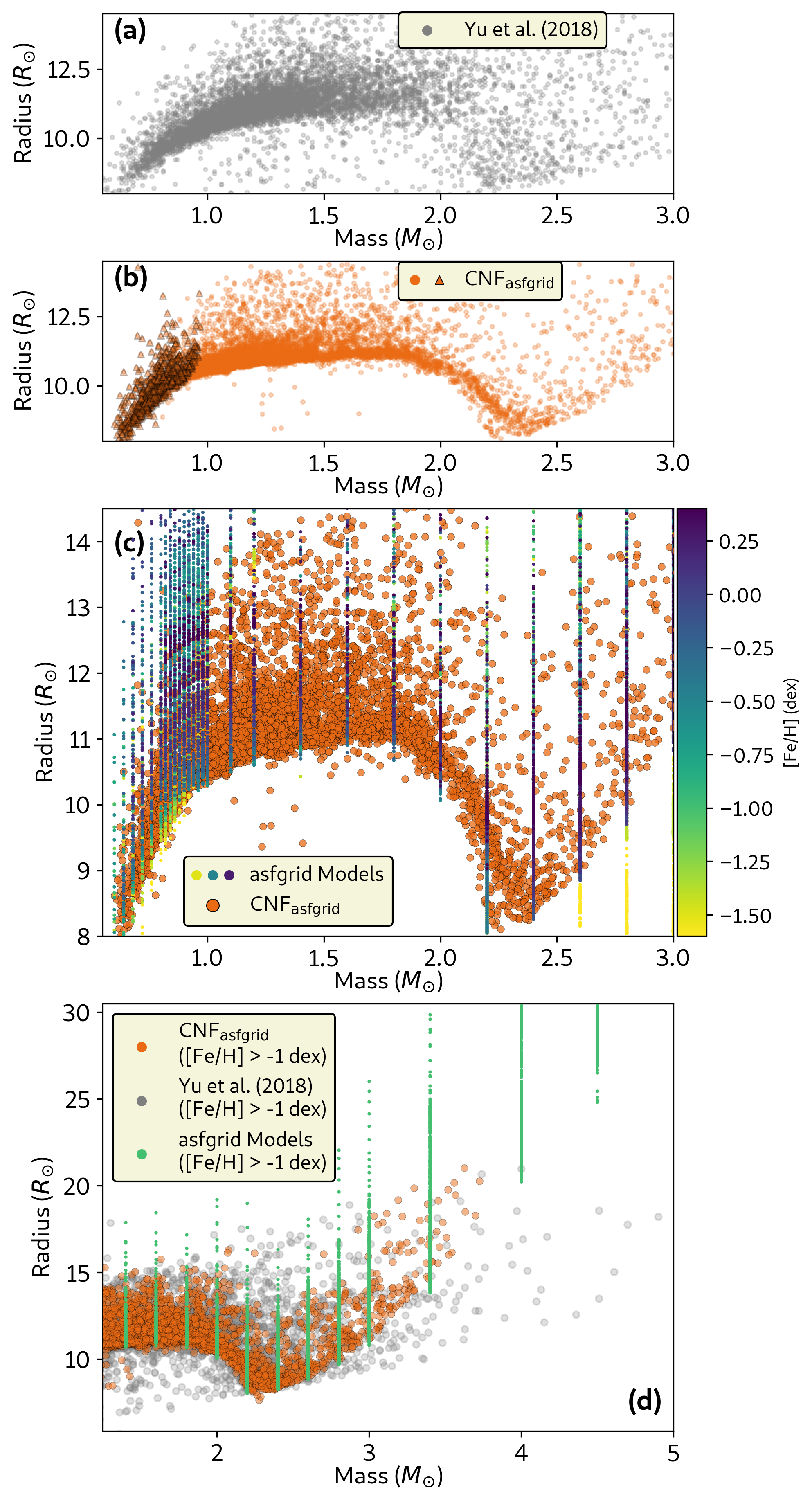}
    \caption{Estimated masses and radii of 7,703 helium core-burning (CHeB) stars from the \citet{Yu_2018} dataset of \textit{Kepler} field red giant stars. \textbf{(a)} Estimates reported by \citet{Yu_2018}, which applied AsfGrid's native interpolation method. \textbf{(b)} Estimates using the median of the posterior distribution of each star from CNF$_{\mathrm{asfgrid}}$. The triangles indicate stars whose model ages are greater than 13.8 Gyr, a heuristic used by \citet{Li_2022_nature} to identify very low-mass red clump stars. \textbf{(c)} 
    The CNF$_{\mathrm{asfgrid}}$ estimates from panel (b) plotted alongside the CHeB models from the AsfGrid's grid, which are colored by their metallicity [Fe/H].  \textbf{(d)} A comparison of estimates between those from CNF$_{\mathrm{asfgrid}}$ and from AsfGrid's native interpolation at the high-mass end for stars with [Fe/H]$>-1\,$dex, which comprises 99\% of the CHeB stars from the \citet{Yu_2018} sample. Also shown are AsfGrid models with the same metallicity range. The \citet{Yu_2018} sample has a lower $\nu_{\mathrm{max}}$ limit of $\sim5\,\mu$Hz, resulting in an upper limit of observed radii of $\sim20\,R_{\odot}$ compared to the grid, which includes models beginning to approach the asymptotic giant branch phase.} 
    \label{fig:zacheb}
\end{figure}

\subsection{Stellar Parameter Inference: Kepler Field Red Giants}\label{section:asfgrid_inference}

We use CNF$_{\mathrm{asfgrid}}$ to estimate the conditional distribution $p(\mathbf{y}|\mathbf{x})$, where $\mathbf{x}=$ $[T_{\mathrm{eff}},$ $\Delta\nu$, $\nu_{\mathrm{max}}$, $R$, $\tau]$ and $\mathbf{y}$ = $[M,$ $[\text{Fe/H}], $ $E]$. Other than $M$,$R$, and $\tau$, all values in these vectors adopt measurements reported by \citet{Yu_2018} for 15,388 \textit{Kepler} red giants. We obtain posterior distributions for $M$ and $R$ of each star by sampling from CNF$_{\mathrm{asfgrid}}$ using a standard Salpeter initial mass function independent of age and metallicity.
Similar to the previous Section, the present-day prior distribution of masses is assumed to follow the initial mass prior distribution, given no mass loss in AsfGrid's grid of models.
Similar to Section \ref{section:open_clusters}, samples are drawn until the discrete distribution $\phi^*$ over $\{\phi_1, \phi_2, \cdots, \phi_n\}$ contains a minimum of 50,000 non-zero importance weights. The reported point estimates for $M$ and $R$ are determined from the median of the posterior distribution based on 50,000 draws from $\phi^*$, with uncertainty intervals defined as the interquartile range between the 25th and 75th percentiles.
We present our new estimates in Table \ref{table:kepler_rg}.

To demonstrate that our inferred parameters are consistent with estimates from other grid-based modelling approaches from literature, we compare our mass and radii estimates with those reported by \citet{Silva-Aguirre_2018} on 1,199 red giant branch stars from the BAyesian STellar Algorithm (BASTA, \citealt{BASTA}). The \citet{Silva-Aguirre_2018} study uses asteroseismic measurements from \citet{Yu_2018}, but adopts spectroscopic measurements $T_{\mathrm{eff}}$ and [Fe/H] from the Data Release 13 of the Sloan Digital Sky Survey \citep{Albareti_2017}. We adopt the \citet{Yu_2018} reported $T_{\mathrm{eff}}$ and [Fe/H] values because we predict $M$ and $R$ over all stars in the \citet{Yu_2018} catalogue instead of only a subset that overlaps with the \citet{Silva-Aguirre_2018} study.

Fig. \ref{fig:asfgrid_basta} shows a comparison of mass and radii estimates, which shows that the CNF$_{\mathrm{asfgrid}}$ results closely match the BASTA values up to a 1-2\% offset in mass and a 1\% offset in radius. These small discrepancies are expected, given differences in the input spectroscopic values used (Fig. \ref{fig:asfgrid_basta} inset). 
Furthermore, both results are based on different evolutionary codes (BASTA by default uses BaSTI isochrones \citep{BASTI}, while AsfGrid uses MESA models), which can lead to slight variations in the output properties predicted across grids (e.g., \citealt{Silva-Aguirre_2020}). Besides a comparison with BASTA, we also include a comparison with the values reported directly from the \citet{Yu_2018} that are based on the asteroseismic scaling relations:
\begin{align}\label{eqn:scaling_relations}
    M &= \bigg( \frac{\nu_{\mathrm{max}}}{\nu_{\mathrm{max,\odot}}}\bigg)^3\bigg(\frac{\Delta\nu}{f_{\Delta\nu}\Delta\nu_{\odot}}\bigg)^{-4}\bigg(\frac{T_{\mathrm{eff}}}{T_{\mathrm{eff, \odot}}}\bigg)^{3/2} \notag \\
      & = \kappa_M \bigg(\frac{T_{\mathrm{eff}}}{T_{\mathrm{eff, \odot}}}\bigg)^{3/2}, \\ 
    R &= \bigg( \frac{\nu_{\mathrm{max}}}{\nu_{\mathrm{max,\odot}}}\bigg)\bigg(\frac{\Delta\nu}{f_{\Delta\nu}\Delta\nu_{\odot}}\bigg)^{-2}\bigg(\frac{T_{\mathrm{eff}}}{T_{\mathrm{eff, \odot}}}\bigg)^{1/2}  \notag \\
      & = \kappa_R \bigg(\frac{T_{\mathrm{eff}}}{T_{\mathrm{eff, \odot}}}\bigg)^{1/2},
\end{align}
where $\nu_{\mathrm{max, \odot}} =3090\mu$Hz, $\Delta\nu_{\odot}=135.1\mu$Hz, and $T_{\mathrm{eff, \odot}} = 5777\,$K. By definition, the `uncorrected' scaling relations have $f_{\Delta\nu} = 1$. Meanwhile, the scaling relations used in the \citet{Yu_2018} work applies $f_{\Delta\nu}$ computed by the AsfGrid code, which are determined by an interpolation over $(\Delta\nu$, $\nu_{\mathrm{max}}$, [Fe/H], $T_{\mathrm{eff}})$ within its accompanying grid \citep{Sharma_2016, Stello_2022}.

Compared to AsfGrid's native interpolated solution, CNF$_{\mathrm{asfgrid}}$'s result in Fig. \ref{fig:asfgrid_basta} shows greater consistency in mass and radii with BASTA (which uses grid-based sampling), as evidenced by its overall smaller fractional deviation and dispersion relative to the BASTA values. This improvement is attributed to the direct sampling of mass and radii values from the emulated grid, which circumvents the need to utilize scaling relations (Eqn. \ref{eqn:scaling_relations}) and their associated correction factors. Importantly, Fig. \ref{fig:asfgrid_basta} also demonstrates that there exists a systematic offset of $\sim8\%$ for mass and $\sim4\%$ for radius between CNF$_{\mathrm{asfgrid}}$ and the \citet{Yu_2018} result for red giant branch stars. 

\begin{figure}
    \centering
	\includegraphics[width=1.\columnwidth]{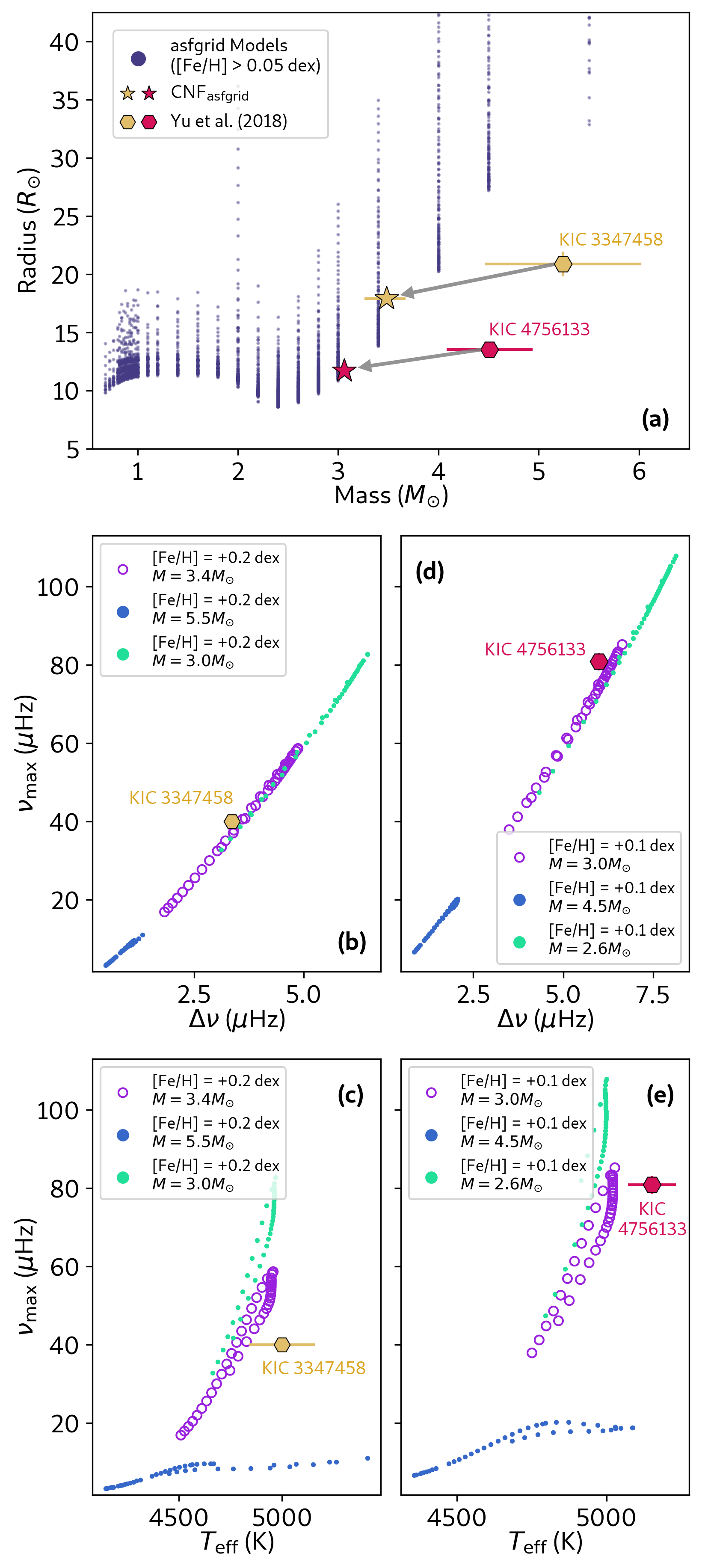}
    \caption{The revision of two high-mass ($M\geq4.5\,M_{\odot}$) helium core-burning stars to lower masses using CNF$_{\mathrm{asfgrid}}$. \textbf{(a)} Shown are new mass and radius estimates of KIC 3347458 ([Fe/H] = $+0.18\pm0.3$ dex) and KIC 4756133 ([Fe/H] = $+0.11\pm0.15$ dex)   from CNF$_{\mathrm{asfgrid}}$, which are now within boundaries of the original grid. The grid does not inherently contain any models with $M\geq4.5\,M_{\odot}$ having $R<25\,R_{\odot}$. \textbf{(b-c)} An examination of models from the grid that have the closest observables ($\nu_{\mathrm{max}}$, $\Delta\nu$, $T_{\mathrm{eff}}$) to those from KIC 3347458. Models from the $M=3.4\,M_{\odot}$ track form the closest match to the observed star. \textbf{(d-e)} The same as panels (b-c), but for KIC 4756133, where models from the $M=3.0\,M_{\odot}$ track form the closest match across all observables.}
    \label{fig:clump_discrepancy}
\end{figure}

\begin{figure}
    \centering
	\includegraphics[width=1.\columnwidth]{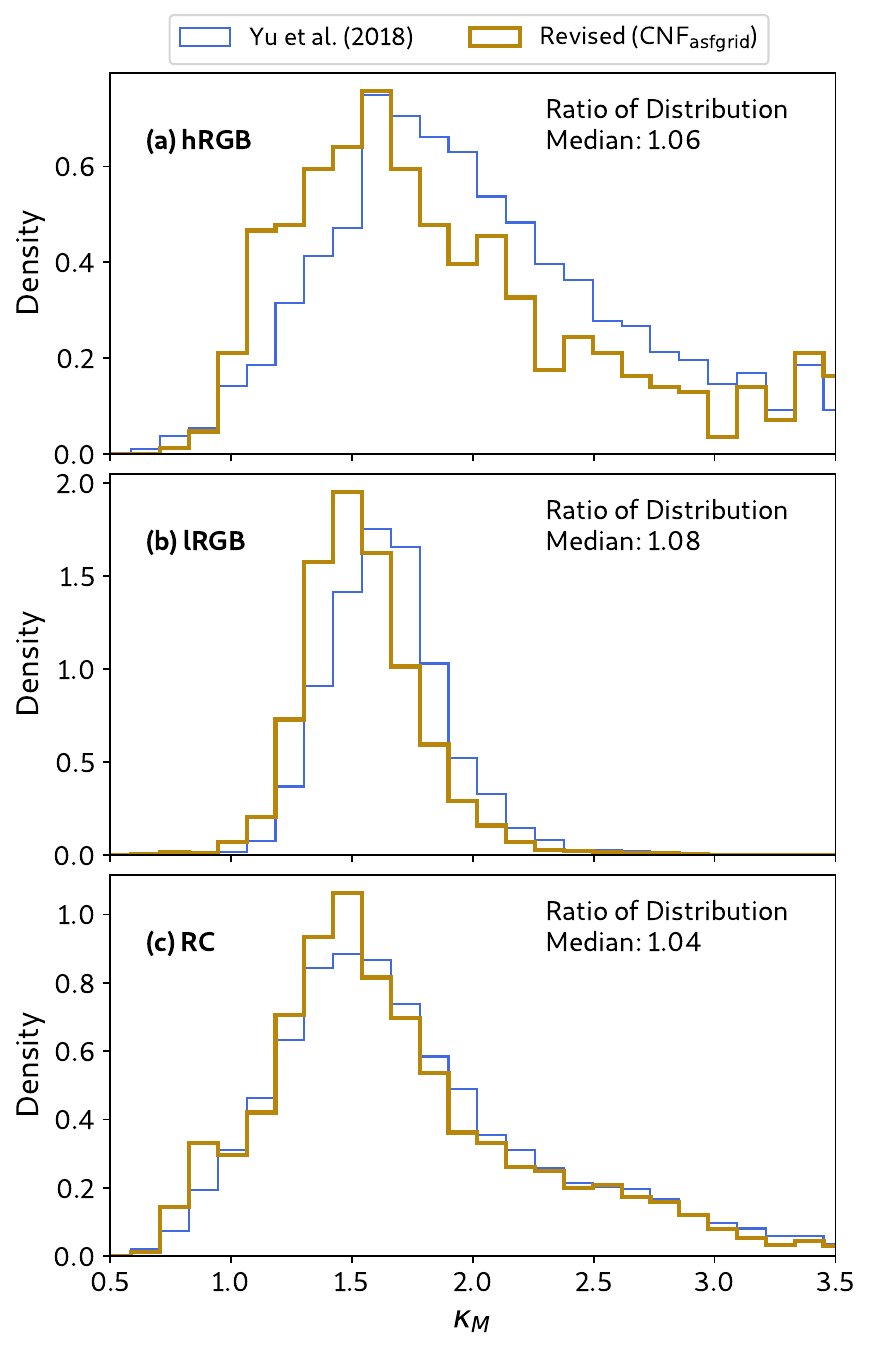}
    \caption{The scaling relation mass coefficient, $\kappa_M$, (Eqn. \ref{eqn:scaling_relations}) of \textit{Kepler} red giants for the CNF$_{\mathrm{asfgrid}}$ estimates in comparison to the \citet{Yu_2018} results. The stars are categorized according to their evolutionary states: (a) high-luminosity hydrogen-shell burning stars (hRGB), (b) low-luminosity hydrogen-shell burning stars (lRGB, (c) helium core-burning stars (RC). Each category follows the definition outlined by \citet{Sharma_2019} (see text). The ratio between the \citet{Yu_2018} median $\kappa_M$ and the revised median $\kappa_M$ from CNF$_{\mathrm{asfgrid}}$ for each evolutionary state is displayed in each panel.}
    \label{fig:kappa_m}
\end{figure}

To examine our results for helium core-burning stars, we compare the masses and radii of 7,703 red clump stars inferred by CNF$_{\mathrm{asfgrid}}$ with the estimates from the \citet{Yu_2018} dataset in Figure \ref{fig:asfgrid_clump_comparison}. Similar to the previous comparison using hydrogen shell-burning stars, the CNF$_{\mathrm{asfgrid}}$ results are systematically smaller by $8-10\%$ in mass and $\sim4\%$ in radius. To investigate the source of this discrepancy, we present these estimates in the mass-radius plane in Fig. \ref{fig:zacheb}. In theory, low-mass red clump stars ignite core helium at similar core masses, which manifest as a sharp lower edge in the mass-radius plane plane corresponding to the zero-age core helium burning stage \citep{Li_2021}. 
Figs. \ref{fig:zacheb}b-c show that the estimates from CNF$_{\mathrm{asfgrid}}$ are strongly consistent with model predictions of the sharp edge, with smoothly interpolated estimates across regions in the grid for which mass is sparsely sampled ($M>1.2M_{\odot}$). Beyond the sharp edge, the CNF$_{\mathrm{asfgrid}}$ estimates for $M\apprge1.8M_{\odot}$ reveal well-defined structures not easily observed in in the \citet{Yu_2018} results (Fig. \ref{fig:zacheb}a). In particular, red clump stars are predicted to transition to lower luminosities and radii as a consequence of diminishing core masses for stars with total masses at and slightly above the helium flash limit at $M\sim1.7-2.0M_{\odot}$ \citep{Girardi_1999}. This transition with mass is expected to be gradual yet distinct. Beyond the helium flash limit, helium core-burning stars are predicted to become larger and more luminous from an increase in helium core mass, leading to the presence of the Vertical Substructure (VS) feature observed in Milky Way red clump populations \citep{Girardi_2016}. The estimates from CNF$_{\mathrm{asfgrid}}$ distinctly show these population-level features as demonstrated in Figs. \ref{fig:zacheb}c-d, consistent with predictions from models.

These new results suggest revisions to earlier estimates of the masses and radii of \textit{Kepler} helium core-burning stars, especially for those at higher masses. Fig. \ref{fig:zacheb}d shows that many stars with a previously estimated $M>3M_{\odot}$ have radii too small relative to the lower boundary of AsfGrid's grid of models in $M-R$ space. According to the grid, such stars are not as massive as previously reported, instead belonging to the VS population representing the transition of helium core-burning stars to higher radii and luminosities. We note that the CNF$_{\mathrm{asfgrid}}$ 
emulation errors (Fig. \ref{fig:asfgrid_validation}) are typically below 3\% in mass and radius for stars with $M>3M_{\odot}$ and [Fe/H] $>$ -1.0 dex, which is much smaller than the reported mass discrepancy of 10-15\%, as shown in Fig. \ref{fig:asfgrid_clump_comparison}b.

To confirm that the CNF$_{\mathrm{asfgrid}}$ estimates are more consistent with Asfgrid's grid compared to the published estimates, we present in Figs. \ref{fig:clump_discrepancy}b-e the observables ($\Delta\nu$, $\nu_{\mathrm{max}}$, $T_{\mathrm{eff}}$) of several evolutionary tracks from the grid that are approximately matched in metallicity to the two \textit{Kepler} red giants but are varied in mass. In both examples, we find that tracks having masses similar to the CNF$_{\mathrm{asfgrid}}$ estimate (lower mass) contain the model forming the closest match of their output properties to the observables of the two red giants in consideration. Meanwhile, tracks having masses similar to the previous estimate (higher mass) do not have any output properties within close proximity to the observables of the two red giants. 
This result verifies the revised lower-mass estimates for these stars from AsfGrid's grid of models, and at the same time indicates that these stars have predicted mean densities that deviate significantly from those expected by the scaling relations. Note that these results only preclude the existence of actual oscillating helium core-burning stars with $M>4\,M_{\odot}$ in the context of AsfGrid's grid of models. Only by further consideration of various input physics from other grids of stellar models alongside more detailed asteroseismic measurements of these stars (e.g., \citealt{Crawford_2024}) can we completely rule out the possibility of very high-mass stars existing in the \textit{Kepler} field.

To summarize our revised estimates of masses and radii for the \textit{Kepler} field red giants, we present in Fig. \ref{fig:kappa_m} a comparison of the quantity $\kappa_M$, which is the seismic component of the mass asteroseismic scaling relation in Eqn. \ref{eqn:scaling_relations}.  The presented histograms of $\kappa_M$ represent the observed mass distribution of \textit{Kepler} red giants within the \textit{Kepler} field, which have persistently been at tension with mass distributions from synthetic populations of the \textit{Kepler} field simulated with the \textit{Galaxia} code \citep{Sharma_2016}. More specifically, \textit{Galaxia} was identified to predict too many low-mass (low $\kappa_M$) stars in comparison to $\kappa_M$ from the \citet{Yu_2018} sample, which were derived using scaling relations and corrected using AsfGrid's native interpolation method. In particular, the ratios between the observed distribution's median $\kappa_M$ with that from \textit{Galaxia} were are about 1.1 for high-luminosity red giant branch stars, 1.09 for low-luminosity red giant branch stars, and 1.02 for helium core-burning stars \citep{Sharma_2019}. These offsets
motivated modifications to the \textit{Galaxia} synthetic model in the form of metal-rich thin and thick discs, such that the $\kappa_M$ distribution of the simulated populations can fit the observed values better. Our revised $\kappa_M$ distributions in Fig. \ref{fig:kappa_m} show qualitative similarity to the original predictions from \textit{Galaxia} (see Fig. 7 of \citealt{Sharma_2016} and Fig. 12 of \citealt{Sharma_2019}), which assumed a relatively metal-poor thick disc and a thin disc whose metallicity decreases with age. Moreover, the derived median ratios of the $\kappa_M$ distribution between the \citet{Yu_2018} sample and our results are quantitatively similar to that between the \citet{Yu_2018} sample and \textit{Galaxia}. This suggests that the 
tension in the synthetic modelling of the \textit{Kepler} field originates not from a Milky Way model mis-specification, but rather from inaccurate mass estimates using AsfGrid's grid of stellar models. In follow-up work we aim to examine if results from CNF$_\mathrm{asfgrid}$ may improve the agreement between observed and synthetic populations in the \textit{K2} fields, which were also studied in the \citet{Sharma_2019} study.

\section{Conclusion}

In this study, we have presented conditional normalizing flows (CNFs) as emulators for grids of stellar evolutionary models. This generative approach is capable of capturing the complex relations between the input parameters $\mathbf{x}$ of a grid with its output properties $\mathbf{y}$. We have trained three CNFs, namely CNF$_{\mathrm{dwarf}}$, CNF$_{\mathrm{giant}}$, and CNF$_{\mathrm{asfgrid}}$, to emulate three grids of models that have multi-dimensional input parameters and predicted output observables. Our main results in this work are the following:

\begin{itemize}
    \item The CNF can efficiently interpolate within the input parameter space, yielding highly flexible conditional distributions $p_N(\mathbf{y}|\mathbf{x})$ that take the form of evolutionary tracks in the emulated grid. Marginalizing across a range of input parameters yields a continuum of evolutionary tracks, which captures the smooth variation of output observables with respect to input parameters that is expected from stellar evolutionary models.
    \item Being able to accept constraints on both $\mathbf{x}$ and $\mathbf{y}$, the CNF is capable of emulating evolutionary trajectories of models having a fixed output parameter, such as isochrones. The CNF presents an instructive tool for visualizing the relations between input and output parameters within the grid. This concept is illustrated through an interactive demonstration of CNF$_{\mathrm{dwarf}}$ presented on an online repository.
    \item We benchmarked the CNFs by comparing the accuracy of their output observables to those from a held-out validation grid. In most regions within the grid, the emulation error incurred is below a fraction of a percent, with exceptions only for extreme cases where models within the original grid are highly sparse or substantially overlapping. 
    \item The CNF can be used as a generative prior for Bayesian inference on a grid of models using Sampling/Importance Resampling. The posterior distribution of grid parameters obtained using this approach is not resolution-limited by the original grid, and contains samples having a one-to-one mapping with samples from the prior distribution and with samples in the output parameter space. We presented the \texttt{modelflows} package \citep{modelflows} to enable stellar parameter inference through this approach using CNF$_{\mathrm{giant}}$ and CNF$_{\mathrm{asfgrid}}$.
    \item Using CNF$_{\mathrm{giant}}$, we inferred stellar parameters for red giants in open clusters NGC 6791 and NGC 6819 using measurements from global asteroseismic parameters and spectroscopy. In both cases, large uncertainties in the age of the cluster were demonstrated as a result of unconstrained priors in initial helium abundances and the value of the mixing length parameter.  
    \item We applied CNF$_{\mathrm{asfgrid}}$ as an emulator to the grid accompanying the AsfGrid code and inferred revised masses and radii for 15,388 \textit{Kepler} red giant stars from the \citet{Yu_2018} sample. Our estimated masses and radii of a subset of red giant branch stars show a robust consistency with BASTA grid-based modelling, with only offsets of $\sim2\%$ in mass and $\sim1\%$ in radius. Meanwhile, our estimates on helium core-burning stars recovered distinct characteristics of secondary clump and vertical substructure populations in mass-radius space, consistent with predictions from stellar models. These new estimates indicate an overestimation of previously reported masses above $3.5\,M_{\odot}$ for helium core-burning stars in the \textit{Kepler} field. More generally, our results suggest that previous applications of the revised asteroseismic scaling relations for \textit{Kepler} red giants have overestimated their masses by $\sim5-10\%$. Our revised mass distribution of \textit{Kepler} red giants indicates that the original \textit{Galaxia} model from the \citet{Sharma_2016} study did not strongly over-predict low-mass red giants as previously inferred.
    \item The trained CNFs, experiments, and interactive visuals in this work are publicly available in the repository\footnote{\url{https://github.com/mtyhon/modelflows}} associated with the \texttt{modelflows} package.
\end{itemize}

The use of generative approaches for statistical inference using simulations is an emerging domain in machine learning. Normalizing flows, especially, have seen use in simulation-based inference tasks \citep{sbi} and even integration in modern sampling techniques such as Markov Chain Monte Carlo methods \citep{Hoffman_2019}. While this study has focused on the use of normalizing flows on grids of evolutionary stellar models, the ability of normalizing flows to interpolate in high dimensions and generative highly expressive conditional distributions is generally valuable across other areas in astronomy in which data can be structured in the form of a high-dimensional grid.

\begin{acknowledgments}
The groundwork for this study was performed at the `Stellar Astrophysics in the Era of Gaia, Spectroscopic, and Asteroseismic Surveys' workshop at the Munich Institute for Astro-, Particle and BioPhysics (MIAPbP) center held during August 2023.
\end{acknowledgments}

\appendix

\section{Masked Autoregressive Flow and Neural Spline Flows}\label{appendix:flows}

The masked autoregressive flow \citep{Papamakarios_2017} learns joint, multi-dimensional densities by considering them as a product of one-dimensional densities, such that the density of the $i$-th variable is conditioned on the $i-1$ variables preceding it. These one-dimensional densities are parameterized as single Gaussians:
\begin{equation}
    p(x_i|\mathbf{x}_{1:i-1}) = \mathcal{N}(x_i| \mu_i, (\text{exp}\,\alpha_i)^2),
\end{equation}
where $\mu_i = f_{\mu}(\mathbf{x}_{1:i-1})$ and $\alpha_i = f_{\alpha}(\mathbf{x}_{1:i-1})$. Here, $f_{\mu}$ and $f_{\alpha}$ are functions that determine the mean and variance of the $i$-th conditional density given the $i-1$ variables. Such functions are implemented with neural networks. Sampling from the autoregressive flow proceeds through recursive sampling across $i$:
\begin{equation}
    x_i = z_i\,\text{exp}\,\alpha_i + \mu_i,
\end{equation}
where $z_i$ is the base distribution, defined as $z_i\sim\mathcal{N}(0,1)$ for one dimensions or a multivariate normal for higher dimensions. By construction, this transformation is easily inverted, such that $z_i = (x_i - \mu_i)\,\text{exp}\,(-\alpha_i)$. In addition, the autoregressive structure of this flow yields a triangular Jacobian for the transformation, such that the logarithm of its absolute determinant is easily computed as $\Sigma_i f_{\alpha_i}(\mathbf{x}_{1:i-1})$.
In practice, the autoregressive property of one-dimensional densities is gained through the use of binary masks within the computational layers of the neural network. Following the construction of the Masked Autoencoder Distribution
Estimator (MADE, \citealt{MADE}), such masks are constructed with a specific ordering to enable individual output nodes of the neural network to predict $p(x_1), p(x_2|x_1), \cdots, p(x_i|\mathbf{x}_{1:i-1})$ within a single pass of data. Therefore, $\mu_i$ and $\alpha_i$ across all $i$ can be output simultaneously from a single network, while maintaining the autoregressive property of the flow.

A neural spline flow  applies the concept of a coupling transform \citep{RealNVP}, namely a transformation that maps an input $\mathbf{z}$ to $\mathbf{x}$ by performing the following operations:
\begin{enumerate}
    \item Split $\mathbf{z}$ into $[\mathbf{z}_{1:i-1}, \mathbf{z}_{i:D}]$, where $D$ is the dimensionality of $\mathbf{z}$.
    \item Compute $x_j = g_{\theta_j}(z_j)$ for $j = i,\cdots, D$ in parallel, where $g_{\mathbf{\theta_j}}$ is a function parameterized by $\mathbf{\theta_i}$ that is learned by a neural network. 
    \item Set $\mathbf{z}_{1:i-1} = \mathbf{x}_{1:i-1}$ and return $\mathbf{x} = [\mathbf{x}_{1:i-1}, \mathbf{x}_{i:D}]$.
\end{enumerate}
Note that following the above approach, vector components $\mathbf{x}_{1:i-1}$ are not directly transformed.
A coupling transform satisfies the conditions for a normalizing flow, in that it has a triangular Jacobian matrix, with the determinant as the product of its diagonal elements.
Constructing a neural spline flow proceeds by specifying $g_{\mathbf{\theta_i}}$ as a monotonic rational quadratic spline function. Briefly, the function splits the input domain into $K$ bins over which the spline is defined. Here, $\mathbf{\theta_i}$ comprises $ = [\mathbf{\theta_i}^w, \mathbf{\theta_i}^h, \mathbf{\theta_i}^d]$, which parameterize the splines by the following:
\begin{itemize}
    \item $\mathbf{\theta_i}^w$ is of length $K$ and determines the width of the bins.
    \item $\mathbf{\theta_i}^h$ is of length $K$ and determines the height of the bins.    
    \item $\mathbf{\theta_i}^d$ is of length $K-1$ and determines the derivatives at the internal knots of the spline.    
\end{itemize}
The coupling is combined with an autoregressive transform by defining splines with parameters $\mathbf{\theta}_{1:i-1}$ that act element-wise on $\mathbf{x}_{1:i-1}$ (such that this part is no longer not transformed), which are learnable by direct optimization using stochastic gradient descent. Meanwhile, $\mathbf{\theta}_{i:D}=\text{NN}(\mathbf{z}_{1:i-1})$, where NN is an autoregressive neural network. Together, these form an Autoregressive Neural Spline Flow (AR-NSF) transform. For further details on the splines, we refer the reader to \citet{Durkan_2019}. 

For CNF$_{\mathrm{dwarf}}$ and CNF$_{\mathrm{giant}}$, we use ten AR-NSF transforms with a ten layer multi-layer perceptron with a width of 256 neurons for their neural networks. For CNF$_{\mathrm{asfgrid}}$, we use eight AR-NSF transforms with a eight layer multi-layer perceptron with a width of 512 neurons for the neural network. We use the default number of spline bins ($K=8$) implemented by \texttt{Zuko} for each transform. The normalizing flows are trained using the Adam optimizer \citep{Adam} with learning rate annealing on plateau, reaching convergence in approximately 1,500-2,000 iterations.

\section{Sobol Sampling Parameter Coverage of MESA Grid of Models} \label{appendix:sobol}

\begin{figure*}
    \centering
	\includegraphics[width=1.\columnwidth]{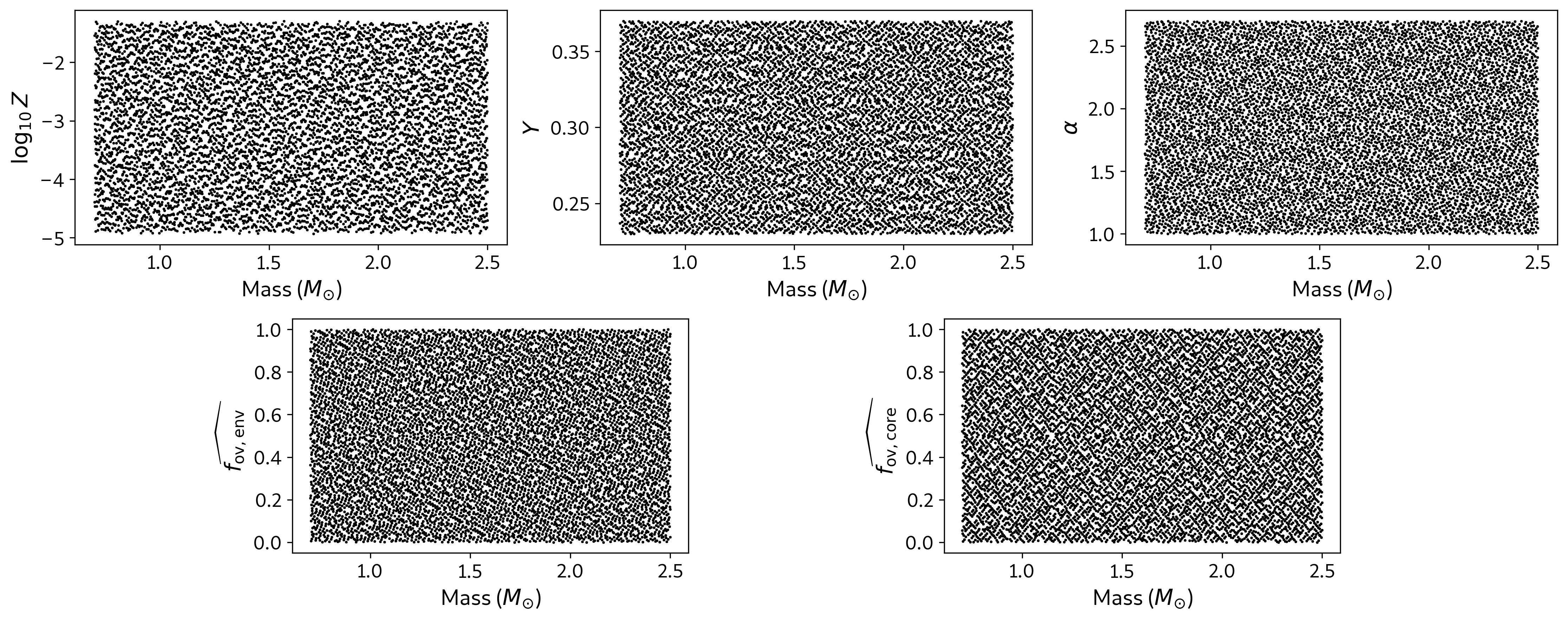}
    \caption{Sobol sequence-sampled points within the six-dimensional parameter space spanned by input parameters $[M, \log_{10}Z, Y, \alpha, \widehat{f_{\mathrm{ov, env}}}, \widehat{f_{\mathrm{ov, core}}} ]$ within the MESA grid of models in theis study. The definitions of each parameter are listed in Table \ref{table:mesa}.}
    \label{fig:sobol}
\end{figure*}

Fig. \ref{fig:sobol} shows the parameter coverage of input parameters in the MESA grid of models (Section \ref{grid_dwarfs}), which form the conditioning variable vector $\mathbf{x}$ to CNF$_{\mathrm{dwarf}}$ and CNF$_{\mathrm{giant}}$.

\section{Interactive Visualization} \label{appendix:interactive}
\begin{figure*}
    \centering
	\includegraphics[width=1.\columnwidth]{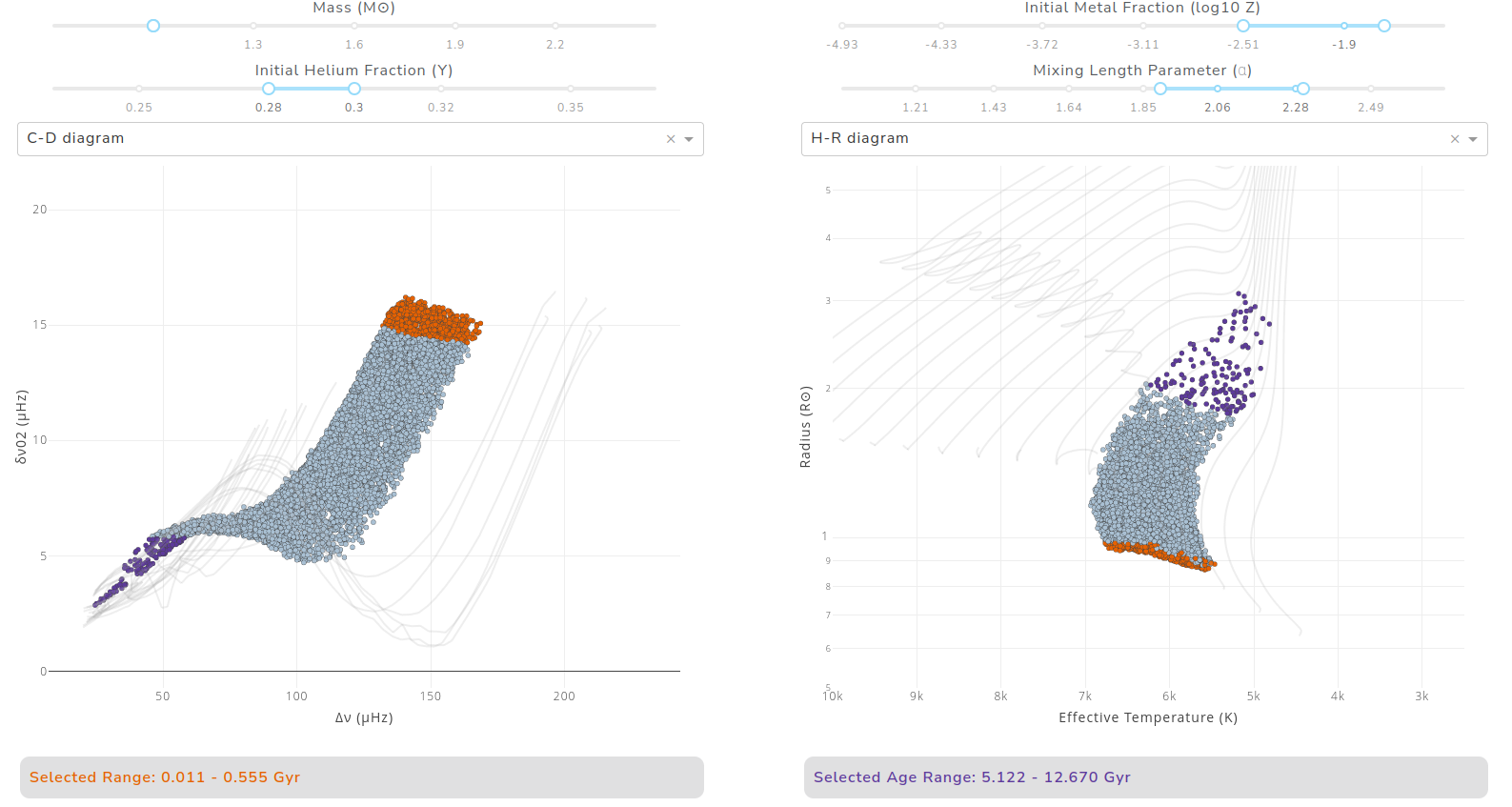}
    \caption{A screenshot of the interactive display tool to visualize evolutionary tracks using CNF$_{\mathrm{dwarf}}$. The points correspond to on-the-fly interpolated samples drawn from CNF$_{\mathrm{dwarf}}$, with sliders at the top to vary input parameters. Shown here are emulated models in the C-D diagram and the H-R diagram for $M=1\,M_{\odot}$, $Y=0.28-0.30$, $\log_{10}Z$=-$2.509$ $-$ -$1.609$, $\alpha=1.9-2.3$. The purple and orange points are user-highlighted samples with a one-to-one correspondence between the two panels. The age range of the two groups of highlighted samples are shown at the bottom. Background tracks are from solar metallicity MIST tracks for the H-R diagram, and near-solar metallicity tracks from the MESA grid in this work for asteroseismic diagrams. }
    \label{fig:modelflows}
\end{figure*}

For demonstrative purposes, we develop an interactive plotting widget using Plotly v2.27 \citep{plotly} that showcases the ability of CNF$_{\mathrm{dwarf}}$ to emulate dwarf and subgiant stellar evolutionary tracks. This source code to run this widget locally is available at \url{https://github.com/mtyhon/modelflows}, with details on an online-only version on the same page.
A screenshot of the widget is shown in Figure \ref{fig:modelflows}, demonstrating samples drawn from CNF$_{\mathrm{dwarf}}$ emulating models across two evolutionary diagrams simultaneously. The widget has the following functionalities:
\begin{itemize}
    \item Interpolation across input parameters $\mathbf{x}=$ [$M$, $Z$, $Y$, $\alpha$] controlled by the user with sliders to provide on-the-fly emulation of models across a range of $\mathbf{x}$.
    \item Visualization of either the C-D diagram, the H-R diagram, or the $\Delta\nu-\epsilon$ diagram. These three diagrams are currently supported at time of release, with more in future releases. Drawn samples from CNF$_{\mathrm{dwarf}}$ are persistent across a change of diagram, which is performed using the drop-down menu below the sliders above each panel.
    \item Highlighting of targets across diagrams by lasso or box select. Samples are colored orange if highlighted in the left panel, while they are colored purple of highlighted in the right panel. The selection of samples in one panel will automatically highlight the same set of samples in the other diagram. This consistency persists even with a change of diagram. The text box at the bottom of each panel displays the age range of each highlighted selection.
\end{itemize}

\bibliography{sample631}{}
\bibliographystyle{aasjournal}

\end{document}